\documentclass[
aps,
prd,
twocolumn,
nofootinbib,
longbibliography
]{revtex4-2}

\usepackage{hyperref}
\usepackage{url}
\usepackage{setspace}
\usepackage{amsmath,amssymb,amscd,braket}
\usepackage{amsfonts}
\usepackage{color}
\usepackage{graphicx}
\usepackage{xcolor}
\usepackage{mathrsfs}
\usepackage{tikz}
\usetikzlibrary{intersections,calc,arrows.meta,decorations.pathmorphing,decorations.markings,shapes.geometric}
\usepackage{caption}
\usepackage{float}

\newcommand{\bea}{\begin{eqnarray}}
\newcommand{\eea}{\end{eqnarray}}
\newcommand{\be}{\begin{equation}}
\newcommand{\ee}{\end{equation}}
\newcommand{\ba}{\begin{align}}
\newcommand{\ea}{\end{align}}

\newcommand{\dd}{\mathrm d}
\newcommand{\ii}{\mathrm i}

\begin{document}
\title{Bouncing and QNM convergence in Kerr
}

\author{Paolo Arnaudo}
\email{p.arnaudo@soton.ac.uk}
\affiliation{Mathematical Sciences and STAG Research Centre, University of Southampton, Highfield, Southampton SO17 1BJ, UK}

\author{Benjamin Withers}
\email{b.s.withers@soton.ac.uk}
\affiliation{Mathematical Sciences and STAG Research Centre, University of Southampton, Highfield, Southampton SO17 1BJ, UK}

\begin{abstract}
We show that the convergence region of the QNM representation of the retarded Green's function of the Kerr spacetime is set by complex-time bouncing singularities, as in our previous work for the Schwarzschild case. The main difference between Kerr and Schwarzschild is that there are two distinct QNM families, prolate and oblate, which have different convergence regions. We provide explicit analytical formulae for the bounce radii and corresponding convergence regions in Boyer-Lindquist coordinates.
We also provide an analysis of convergence in the Reissner-Nordstr\"om spacetime.
\end{abstract}

\maketitle
\tableofcontents

\section{Introduction}

The study of black hole geometries and their linear perturbations is of primary interest in this age of increasingly accurate theoretical predictions of general relativity and a growing number of gravitational-wave observations \cite{LVKloud, Berti:2025hly}.
Relevant properties of the underlying geometry are indeed encoded in the response of the system when being perturbed. At linear order, a key object to analyse the system is the retarded Green's function. Based on the seminal works \cite{Leaver, Andersson:1996cm}, recent developments have shown that the retarded Green's function cannot be represented as a quasinormal mode (QNM) sum everywhere in spacetime \cite{Kuntz:2025gdq, Arnaudo:2025uos}.

At fixed spin and angular quantum numbers $(s,\ell,m)$, an important question is where the QNM contribution to the retarded Green's function converges for given source and observation points. This question depends on the large-overtone behaviour of both the frequencies and their residues, and connects the spectral properties of perturbations to the geometry of the background.

For Schwarzschild, the geometric origin of this convergence region was identified in \cite{Arnaudo:2026tcy}. The relevant singularities occur at complex times and are associated with bouncing null curves, obtained as limits of spacelike geodesics connecting the two exterior regions of the maximally extended spacetime. This mechanism determines a bounce radius $r^{\mathrm{bounce}}\simeq 2.556929M$, bounding the QNM convergence region. 

The realisation of this picture originates from the AdS/CFT literature \cite{Fidkowski:2003nf, Festuccia:2005pi} and received much attention in recent developments \cite{Parisini:2023nbd,Ceplak:2024bja,Buric:2025anb,Buric:2025fye,Afkhami-Jeddi:2025wra,Dodelson:2025jff,Ceplak:2025dds,Jia:2025jbi,Giombi:2026kdz,Arnaudo:2026der,Grozdanov:2026cut,Jia:2026ryl,Grozdanov:2026ktq,Arnaudo:2026axe, Barrat:2026jfg}.

The presence of rotation makes the identification more involved due to the richer structure of the problem. In Kerr spacetime, the spherical symmetry of Schwarzschild is reduced to axial symmetry, and the angular and radial perturbation equations are coupled through a frequency-dependent separation constant. Consequently, even at fixed $(s,\ell,m)$, the angular asymptotics enter the geometry governing large-frequency radial propagation. Moreover, Kerr black holes possess an inner horizon, which changes the maximal extension and introduces an additional horizon logarithm in the radial optical coordinates. The global continuation relevant to the QNM problem therefore has a more intricate structure.

As discussed in \cite{Arnaudo:2026tcy}, the problem of identifying the region of convergence of QNMs at fixed quantum numbers $\ell,m$ corresponds to finding the appropriate singularities of the Green's function, which may be in general complex. These singularities can be understood as arising from the initial insertion at $t', r'$, as they are propagated as null rays through a 2$D$ optical metric. In the Schwarzschild case studied in \cite{Arnaudo:2026tcy}, this propagation is governed precisely by radial null geodesics of the Schwarzschild spacetime. In the present case, however, the coupling with the angular problem makes the analysis richer. Here, the appropriate 2$D$ optical metric itself depends on the large-$\omega$ behaviour of the separation constant $\lambda$, and so of the angular eigenvalue $A$.

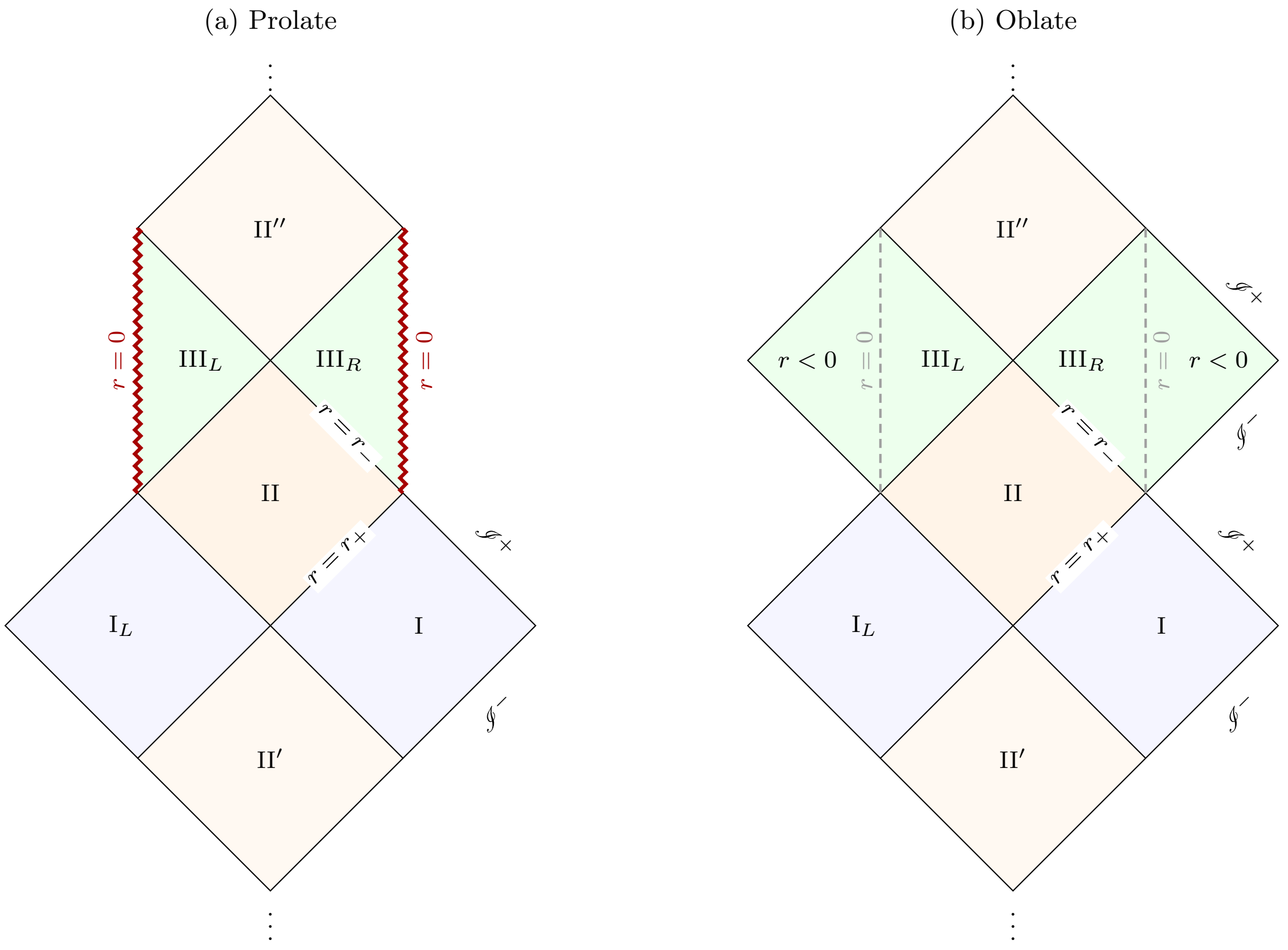
\begin{figure*}[t]
\centering
\begin{minipage}{0.85\textwidth}
\centering
\resizebox{\linewidth}{!}{%
\begin{tikzpicture}[x=1.55cm,y=1.55cm,every node/.style={font=\footnotesize}]
\foreach \panelshift/\panel in {0/0,5.6/1} {
\begin{scope}[shift={(\panelshift,0)}]

  \fill[blue!4] (0,0)--(1,1)--(2,0)--(1,-1)--cycle;
  \fill[blue!4] (0,0)--(-1,1)--(-2,0)--(-1,-1)--cycle;
  \fill[orange!9] (0,0)--(1,1)--(0,2)--(-1,1)--cycle;
  \fill[orange!5] (0,0)--(1,-1)--(0,-2)--(-1,-1)--cycle;
  \fill[orange!5] (0,2)--(1,3)--(0,4)--(-1,3)--cycle;
  \fill[green!7] (-1,1)--(0,2)--(-1,3)--cycle;
  \fill[green!7] (1,1)--(0,2)--(1,3)--cycle;

  \draw (0,0)--(1,1)--(2,0)--(1,-1)--cycle;
  \draw (0,0)--(-1,1)--(-2,0)--(-1,-1)--cycle;
  \draw (1,1)--(0,2)--(-1,1);
  \draw (1,-1)--(0,-2)--(-1,-1);
  \draw (0,2)--(1,3)--(0,4)--(-1,3)--cycle;

  \ifnum\panel=0\relax
    \node[font=\small] at (0,4.55) {(a) Prolate};
    \draw[very thick,red!65!black,decorate,decoration={zigzag,amplitude=1pt,segment length=4pt}] (-1,1)--(-1,3);
    \draw[very thick,red!65!black,decorate,decoration={zigzag,amplitude=1pt,segment length=4pt}] (1,1)--(1,3);
    \node[red!65!black,rotate=90] at (-1.16,2) {$r=0$};
    \node[red!65!black,rotate=90] at (1.16,2) {$r=0$};
  \else
    \node[font=\small] at (0,4.55) {(b) Oblate};
    \fill[green!7] (-1,1)--(-2,2)--(-1,3)--cycle;
    \fill[green!7] (1,1)--(2,2)--(1,3)--cycle;
    \draw (-1,1)--(-2,2)--(-1,3);
    \draw (1,1)--(2,2)--(1,3);
    \draw[gray!75,thick,densely dashed] (-1,1)--(-1,3);
    \draw[gray!75,thick,densely dashed] (1,1)--(1,3);
    \node[gray!75,rotate=90] at (-1.12,2) {$r=0$};
    \node[gray!75,rotate=90] at (1.12,2) {$r=0$};
    \node at (-1.55,2) {$r<0$};
    \node at (1.55,2) {$r<0$};
    \node[rotate=-45] at (1.76,2.52) {$\mathscr I^+$};
    \node[rotate=45] at (1.76,1.48) {$\mathscr I^-$};
  \fi

  \node at (1.12,0) {$\mathrm I$};
  \node at (-1.12,0) {$\mathrm I_L$};
  \node at (0,1) {$\mathrm{II}$};
  \node at (0,-1) {$\mathrm{II}'$};
  \node at (-.52,2) {$\mathrm{III}_L$};
  \node at (.52,2) {$\mathrm{III}_R$};
  \node at (0,3) {$\mathrm{II}''$};
  \node at (0,4.20) {$\vdots$};
  \node at (0,-2.20) {$\vdots$};

  \node[rotate=45,fill=white,inner sep=1pt] at (.52,.52) {$r=r_+$};
  \node[rotate=-45,fill=white,inner sep=1pt] at (.57,1.43) {$r=r_-$};
  \node[rotate=-45] at (1.70,.65) {$\mathscr I^+$};
  \node[rotate=45] at (1.70,-.65) {$\mathscr I^-$};

\end{scope}
}
\end{tikzpicture}%
}
\captionsetup{width=\linewidth,justification=justified,singlelinecheck=false}
\caption{Conformal diagrams of the real sections of the prolate (left) and oblate (right) optical geometries, for $0<a<M$. Regions I are exteriors with $r>r_+$, regions II lie between the horizons, and regions III lie inside the inner horizon. In the prolate geometry, $r=0$ is a timelike singular boundary (red zigzag). In the oblate geometry, $r=0$ is regular (gray dashed), and the extension reaches negative $r$ and additional null infinities. The ellipses indicate further copies of the horizon regions.}
\label{fig:prolate-oblate-penrose}
\end{minipage}
\end{figure*}

In particular, two possible situations arise, which were already studied in the literature \cite{Berti:2005gp, Keshet:2007nv, Keshet:2012uq}.
The first case is the \emph{prolate} sector, in which $A$ grows linearly with $\omega$, and therefore $\lambda\sim a^2\omega^2$ \cite{Keshet:2007nv}. We show that this gives rise to an optical metric
\begin{equation}\label{eq:metric-2D}
\dd s^2_{\text{pr}}=-\frac{\Delta}{\sqrt{(r^2+a^2)^2-a^2\Delta}}\,\dd t^2+\frac{\sqrt{(r^2+a^2)^2-a^2\Delta}}{\Delta}\,\dd r^2.
\end{equation}
The second case is the \emph{oblate} asymptotic sector, where the angular eigenvalues grow as $A\sim-a^2\omega^2$, and the corresponding optical metric is
\begin{equation}
\dd s^2_{\mathrm{ob}}=-\frac{\Delta}{r^2+a^2}\dd t^2+\frac{r^2+a^2}{\Delta}\dd r^2.
\end{equation}
Each of these cases is associated with a different Penrose diagram, as illustrated in Figure \ref{fig:prolate-oblate-penrose}.

The result of this analysis will be the identification of \emph{two} real bounce radii $r^\text{bounce}_\text{prolate}$ and $r^\text{bounce}_\text{oblate}$, which we give precise values for below. These independently determine the convergence regions of the prolate and oblate QNM families respectively, and come from singularities in the complex $t$ plane, associated with bouncing geodesics of the 2D optical metric. In particular, they delineate the regions of QNM convergence by null rays propagating in their respective optical metrics.

\begin{figure}[h!]
\centering
\begin{tikzpicture}[x=0.8*\columnwidth,y=4.5*\columnwidth,every node/.style={font=\normalsize}]
\definecolor{bounceblue}{RGB}{36,96,170}
\definecolor{bouncered}{RGB}{187,60,60}

\foreach \xx/\lab in {0/0,.2/0.2,.4/0.4,.6/0.6,.8/0.8,1/1} {
  \draw[gray!20] (\xx,2.405)--(\xx,2.572);
  \draw (\xx,2.405)--(\xx,2.403);
  \node[below] at (\xx,2.402) {$\lab$};
}
\foreach \yy in {2.42,2.44,2.46,2.48,2.50,2.52,2.54,2.56} {
  \draw[gray!20] (0,\yy)--(1,\yy);
  \draw (0,\yy)--(-.006,\yy);
  \node[left] at (-.008,\yy) {$\yy$};
}
\draw (0,2.405) rectangle (1,2.572);
\node at (.5,2.386) {$a/M$};
\node[rotate=90] at (-.18,2.489) {$r^{\mathrm{bounce}}/M$};

\draw[gray!65,densely dashed] (0,2.556929086)--(1,2.556929086);

\draw[bounceblue,line width=1.1pt] plot coordinates {
(0.00000,2.556929086) (0.00100,2.556938419) (0.00500,2.557009942)
(0.01000,2.557132366) (0.02000,2.557431797) (0.03000,2.557770063)
(0.04000,2.558127015) (0.06000,2.558853353) (0.08000,2.559551768)
(0.10000,2.560187307) (0.12000,2.560736084) (0.14000,2.561180504)
(0.16000,2.561506977) (0.18000,2.561704669) (0.20000,2.561764730)
(0.22000,2.561679799) (0.24000,2.561443660) (0.26000,2.561050997)
(0.28000,2.560497211) (0.30000,2.559778286) (0.32000,2.558890681)
(0.34000,2.557831249) (0.36000,2.556597165) (0.38000,2.555185875)
(0.40000,2.553595052) (0.42000,2.551822555) (0.44000,2.549866403)
(0.46000,2.547724745) (0.48000,2.545395837) (0.50000,2.542878029)
(0.52000,2.540169738) (0.54000,2.537269443) (0.56000,2.534175668)
(0.58000,2.530886971) (0.60000,2.527401936) (0.62000,2.523719159)
(0.64000,2.519837248) (0.66000,2.515754809) (0.68000,2.511470444)
(0.70000,2.506982740) (0.72000,2.502290269) (0.74000,2.497391579)
(0.76000,2.492285191) (0.78000,2.486969595) (0.80000,2.481443242)
(0.82000,2.475704544) (0.84000,2.469751868) (0.86000,2.463583534)
(0.88000,2.457197806) (0.90000,2.450592896) (0.92000,2.443766954)
(0.94000,2.436718066) (0.95000,2.433109401) (0.96000,2.429444251)
(0.97500,2.423840052) (0.98000,2.421943459) (0.99000,2.418107285)
(0.99900,2.414605532) (0.99990,2.414252785) (1.00000,2.414213562)
};

\draw[bouncered,line width=1.1pt] plot coordinates {
(0.00000,2.556929086) (0.00100,2.556929634) (0.00500,2.556938410)
(0.01000,2.556958838) (0.02000,2.557017964) (0.03000,2.557089520)
(0.04000,2.557164611) (0.06000,2.557302465) (0.08000,2.557397497)
(0.10000,2.557427039) (0.12000,2.557374505) (0.14000,2.557227123)
(0.16000,2.556974738) (0.18000,2.556609098) (0.20000,2.556123386)
(0.22000,2.555511907) (0.24000,2.554769854) (0.26000,2.553893147)
(0.28000,2.552878298) (0.30000,2.551722315) (0.32000,2.550422622)
(0.34000,2.548976997) (0.36000,2.547383518) (0.38000,2.545640523)
(0.40000,2.543746576) (0.42000,2.541700432) (0.44000,2.539501020)
(0.46000,2.537147417) (0.48000,2.534638830) (0.50000,2.531974586)
(0.52000,2.529154113) (0.54000,2.526176931) (0.56000,2.523042643)
(0.58000,2.519750926) (0.60000,2.516301525) (0.62000,2.512694244)
(0.64000,2.508928943) (0.66000,2.505005532) (0.68000,2.500923967)
(0.70000,2.496684248) (0.72000,2.492286412) (0.74000,2.487730537)
(0.76000,2.483016730) (0.78000,2.478145137) (0.80000,2.473115931)
(0.82000,2.467929315) (0.84000,2.462585523) (0.86000,2.457084813)
(0.88000,2.451427474) (0.90000,2.445613819) (0.92000,2.439644186)
(0.94000,2.433518941) (0.95000,2.430398085) (0.96000,2.427238474)
(0.97500,2.422426508) (0.98000,2.420803200) (0.99000,2.417527648)
(0.99900,2.414546703) (0.99990,2.414246894) (1.00000,2.414213562)
};

\fill[black] (0,2.556929086) circle (1.7pt);
\draw[black,fill=white,line width=.8pt] (1,2.414213562) circle (2pt);

\draw[bounceblue,line width=1.1pt] (.08,2.452)--(.15,2.452);
\node[anchor=west,bounceblue] at (.17,2.452) {Prolate};
\draw[bouncered,line width=1.1pt] (.08,2.440)--(.15,2.440);
\node[anchor=west,bouncered] at (.17,2.440) {Oblate};

\end{tikzpicture}
\caption{Optical bounce radii which determine QNM convergence in Kerr, as functions of the rotation parameter. The red curve is the oblate bounce, defined by \eqref{oblatebounce}, and the blue curve is the prolate bounce, defined by \eqref{prolatebounce}. The dashed horizontal line marks their common Schwarzschild value $r^{\mathrm{bounce}}/M=2(1+W_0(e^{-1}))$ $\simeq 2.556929$. The open endpoint denotes their common limit $r^{\mathrm{bounce}}/M\to 1+\sqrt{2}$ as $a/M\to1^-$.}
\label{fig:kerr-bounce-radii}
\end{figure}
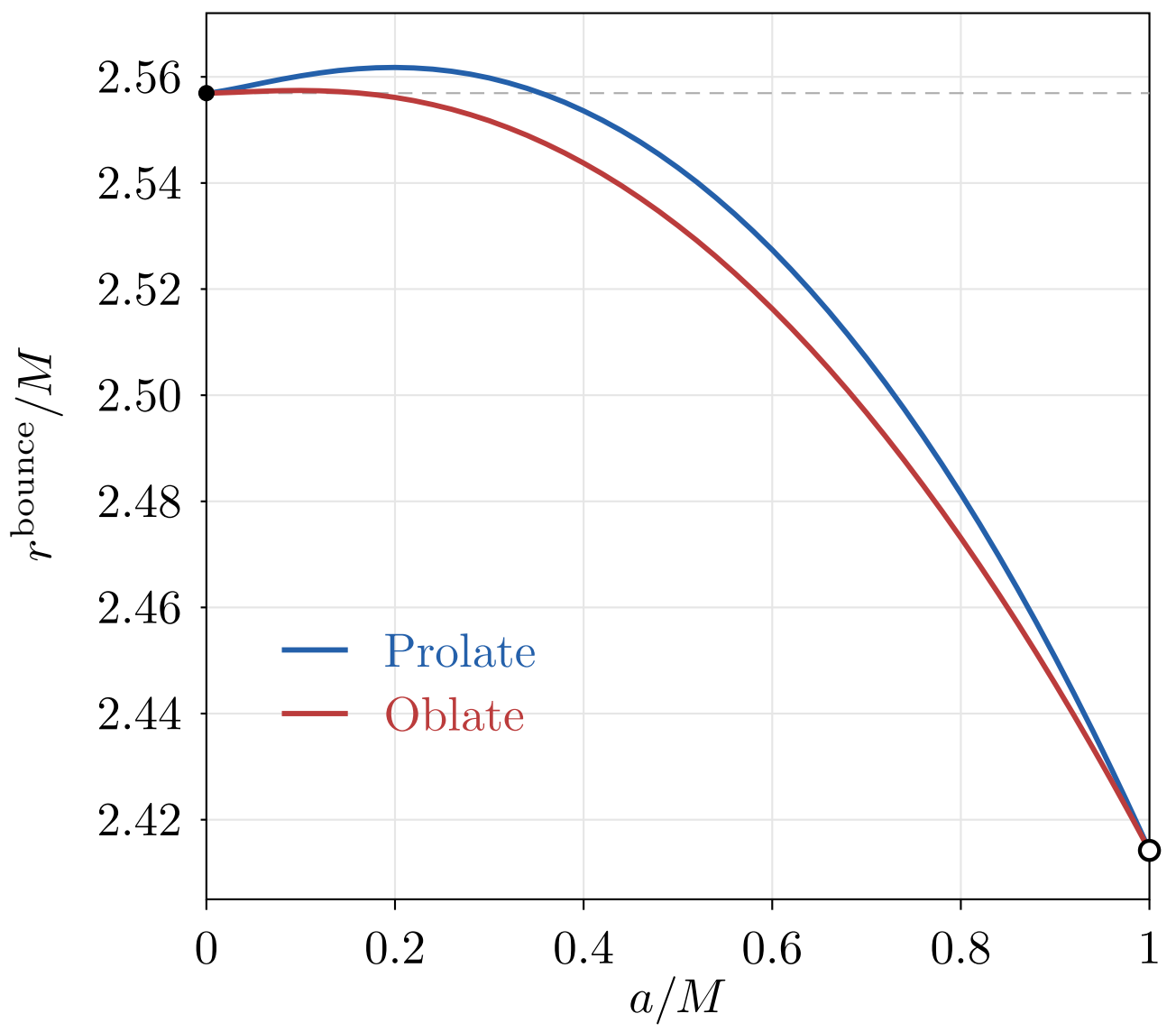

We now summarise the result for these bounce radii, before going into detail in the following sections. In the oblate family the bounce radius $r^\text{bounce}_\text{oblate}$ is given by the relation 
\be
r_*(r^{\mathrm{bounce}}_\text{oblate})=\mathrm{Re}\,r_*(\ii a) \label{oblatebounce}
\ee
where $a$ is the rotation parameter, 
\begin{equation}\label{eq:numerical-optical-normalisation}
r_*(r)=r+\frac{2Mr_+}{r_+-r_-}\log\frac{r-r_+}{2M}-\frac{2Mr_-}{r_+-r_-}\log\frac{r-r_-}{2M},
\end{equation}
and $r_{\pm}$ are the outer and inner horizons. In the prolate family the bounce radius $r^\text{bounce}_\text{prolate}$ is given by the relation
\be
x(r^{\mathrm{bounce}}_\text{prolate})=\mathrm{Re}\,x(r_c) \label{prolatebounce}
\ee
where $x$ is the tortoise coordinate for the 2D prolate optical metric,
\begin{equation}\label{eq:intro2DKerr-tortoise}
\frac{\dd x}{\dd r}=\frac{\sqrt{(r^2+a^2)^2-a^2\Delta}}{\Delta},
\end{equation}
with the integration constant fixed in such a way that $x-r_*\to0$ at infinity. The parameter $r_c$ is the complex root with $\mathrm{Re}\,r_c>0$ and $\mathrm{Im}\,r_c>0$, of the cubic $r^3+a^2r+2Ma^2=0$, explicitly,
\begin{equation}\label{eq:intro-rc}
\begin{aligned}
r_c&=\frac{r_1-r_2}{2}+\ii\,\frac{\sqrt3}{2}\,(r_1+r_2),\\
r_1,r_2&=\left(\sqrt{M^2a^4+\frac{a^6}{27}}\pm Ma^2\right)^{1/3}.
\end{aligned}
\end{equation}
Both bounce radii can be easily obtained numerically using the above formulae and we show the result in Figure \ref{fig:kerr-bounce-radii}. 

We reiterate that the correct null rays which bounce from these locations are those associated to the corresponding 2D optical metric for each family. We provide a full specification of the convergence regions in Section \ref{sec:combined}.

The paper is organised as follows. In Section \ref{sec:setup} we first introduce the Kerr perturbation equations and the connection coefficients entering the radial Green's function. In Sections \ref{sec:prolate} and \ref{sec:oblate} we analyse the prolate and oblate QNM families, respectively, relating their complex bouncing singularities to the corresponding convergence regions and comparing the geometric predictions with numerical results for the frequencies and residues. In Section \ref{sec:combined} we combine the two convergence regions and discuss the resulting spacetime decomposition. Finally, in the discussion, Section \ref{sec:discussion}, we examine the Schwarzschild limit and compare our findings with the Reissner--Nordstr\"om case.

\section{Setup} \label{sec:setup}
The Kerr metric in Boyer-Lindquist coordinates is
\begin{equation}\label{BL}
\begin{aligned}
\dd s^2=-&\frac{\Delta}{\Sigma}(\dd t-a\sin^2\theta\,\dd\phi)^2+\frac{\Sigma}{\Delta}\dd r^2+\\
&\Sigma\,\dd\theta^2+\frac{\sin^2\theta}{\Sigma}\left[\left(r^2+a^2\right)\dd\phi-a\,\dd t\right]^2,
\end{aligned}
\end{equation}
where
\begin{equation}
\begin{aligned} \label{sigmadelta}
\Sigma&=r^2+a^2\cos^2\theta,\\
\Delta&=r^2-2Mr+a^2=(r-r_+)(r-r_-), 
\end{aligned}
\end{equation}
with $r_\pm=M\pm\sqrt{M^2-a^2}$ being the outer and inner horizons.
The surface gravities and angular velocities at the outer and inner horizons read
\begin{equation}\label{eq:kappa-and-Omega}
\kappa_\pm=\frac{r_+-r_-}{4Mr_\pm}>0,\qquad
\Omega_\pm=\frac{a}{2Mr_\pm}.
\end{equation}

Based on the Newman-Penrose formalism, Teukolsky \cite{PhysRevLett.29.1114} showed that in terms of curvature invariants, the perturbation equations decouple and separate.
We consider the Fourier-transform of a spin-$s$ field $\Psi_s(t,r,\theta,\phi)$ and expand it in spin-weighted spheroidal harmonics as
\begin{equation}
\Psi_s=\frac{1}{2\pi}\int e^{-i\,\omega\,t}\sum_{\ell=|s|}^{\infty}\sum_{m=-\ell}^{\ell}e^{i\,m\,\phi}{}_sS_{\ell\,m}(\theta;a\omega)R_{\ell\,m}(r;\omega)\,\mathrm{d}\omega,
\end{equation}
where $\omega$ is the frequency of the perturbation, and $\ell$ and $m$ are the angular quantum numbers.

The perturbation equation then reduces to a system of ODEs, the angular and radial ones, coupled through a separation constant defined in terms of eigenvalues of the spheroidal harmonics.

By taking this into account, the QNM problem can be studied by focusing on studying the radial equation, which reads
\begin{equation}\label{eq:radial}
\Delta^{-s}\frac{\dd}{\dd r}\left(\Delta^{s+1}\frac{\dd R}{\dd r}\right)
+\left[\frac{K^2-2\ii s(r-M)K}{\Delta}+4\ii s\omega r-\lambda\right]R=0,
\end{equation}
where $K=\left(r^2+a^2\right)\omega-am$ and $\lambda=A+a^2\omega^2-2am\omega$, $A=A_{s\ell m}(a\omega)$ being the angular eigenvalues.

The QNM frequencies are selected by imposing an ingoing boundary condition at the outer horizon $r=r_+$ and an outgoing boundary condition as $r\to\infty$. Explicitly, these read\footnote{Here and throughout this paper $\sim$ means asymptotic to.}
\begin{equation}\label{BC}
\begin{aligned}
R&\sim\Delta^{-s}e^{-\ii(\omega-m\Omega_+)r_*}\qquad &r\to r_+,\\
R&\sim r^{-1-2s}e^{+\ii\omega r_*}\qquad &r\to\infty.
\end{aligned}
\end{equation}

Let us define a new wave function
\begin{equation}
\psi=\Delta^{(s+1)/2}R,
\end{equation}
and let $P=(s+1)\frac{\Delta'}{\Delta}$. In this way, the radial ODE can be written in normal form,
\begin{equation}\label{eq:normalform}
\begin{aligned}
&\psi''+V\,\psi=0,\\
&V=\frac{K^2-2\ii s(r-M)K}{\Delta^2}
+\frac{4\ii s\omega r-\lambda}{\Delta}-\frac{P'}2-\frac{P^2}4.
\end{aligned}
\end{equation}

At the outer horizon, the normalisations of the local radial solutions are
\begin{equation}\label{eq:horizon-basis}
R_{\mathrm{in}}\sim\Delta^2e^{-\ii(\omega-m\Omega_+)r_*},\qquad R_{\mathrm{out}}\sim e^{\ii(\omega-m\Omega_+)r_*},
\end{equation}
and at infinity they are
\begin{equation}\label{eq:infinity-basis}
R_{\mathrm{up}}\sim r^3e^{\ii\omega r_*},\qquad R_{\mathrm{down}}\sim r^{-1}e^{-\ii\omega r_*}.
\end{equation}
The connection coefficients are defined by
\begin{equation}\label{eq:connection-definitions}
\begin{aligned}
R_{\mathrm{in}}&=C_{\mathrm{in},\mathrm{up}}R_{\mathrm{up}}+C_{\mathrm{in},\mathrm{down}}R_{\mathrm{down}},\\
R_{\mathrm{up}}&=C_{\mathrm{up},\mathrm{in}}R_{\mathrm{in}}+C_{\mathrm{up},\mathrm{out}}R_{\mathrm{out}}.
\end{aligned}
\end{equation}
We use the constant weighted Wronskian
\begin{equation}\label{eq:latest-W-definition}
\mathcal W[R_a,R_b]=\Delta^{-1}\left(R_b\frac{\dd R_a}{\dd r}-R_a\frac{\dd R_b}{\dd r}\right),
\end{equation}
and denote $W\equiv \mathcal W[R_{\mathrm{in}},R_{\mathrm{up}}]$.
The above normalisations imply
\begin{equation}\label{eq:basis-W}
\begin{aligned}
&\mathcal W[R_{\mathrm{down}},R_{\mathrm{up}}]=-2\ii\omega,\\ 
&\mathcal W[R_{\mathrm{in}},R_{\mathrm{out}}]=2(r_+-r_-)-2\ii(\omega-m\Omega_+)(r_+^2+a^2).
\end{aligned}
\end{equation}

The three equivalent expressions for a QNM residue are
\begin{equation}\label{eq:three-residues}
\begin{aligned}\underset{\omega=\omega_n}{\mathrm{Res}}\widetilde G&=-\frac{1}{2\ii\omega_n}\underset{\omega=\omega_n}{\mathrm{Res}}\frac{C_{\mathrm{in},\mathrm{up}}}{C_{\mathrm{in},\mathrm{down}}}R_{\mathrm{up},n}(r)R_{\mathrm{up},n}(r')\\&=\frac{R_{\mathrm{up},n}(r)R_{\mathrm{in},n}(r')}{\left.\frac{\dd W}{\dd\omega}\right|_{\omega_n}}\\&=\frac{C_{\mathrm{up},\mathrm{in}}(\omega_n)R_{\mathrm{in},n}(r)R_{\mathrm{in},n}(r')}{\left.\frac{\dd W}{\dd\omega}\right|_{\omega_n}}.\end{aligned}
\end{equation}
The three above forms will be used interchangeably, depending on the radial locations of $r,r'$.

The QNM contribution to the Green's function is
\begin{equation}\label{eq:radial-sum}
G^{\mathrm{QNM}}(t-t';r,r')=-\ii\sum_n e^{-\ii\omega_n (t-t')}\underset{\omega=\omega_n}{\mathrm{Res}}\widetilde G.
\end{equation}
We retain only the leading large-$n$ behaviour in the sums below. Constants may depend on $a,r,r'$ but not on $n$ or $t-t'$, and overall constant phases are suppressed.

In the next two sections, we analyse the two sectors separately. For concreteness, we set $M=1$ and take $\ell=m=2$ and $s=-2$. In each case, we analytically compute the location of candidate bouncing singularities responsible for the finite region of convergence of the QNM sums, and confirm by showing that the large-order sums of QNMs exhibit singularities at a subset of these locations.

\section{Prolate QNM family} \label{sec:prolate}

The first case we analyse is the \emph{prolate} sector, in which $A$ grows linearly with $\omega$, and therefore $\lambda\sim a^2\omega^2$ \cite{Keshet:2007nv}. In this case, the large-$\omega$ expansion of the potential $V$ in \eqref{eq:normalform} reads
\begin{equation}
V=\omega^2\frac{(r^2+a^2)^2-a^2\Delta}{\Delta^2}+\mathcal O(\omega).
\end{equation}
We can identify a two-dimensional metric by requiring that its null curves reproduce the characteristics
\begin{equation}\label{eq:radial-characteristics}
\frac{\dd t}{\dd r}=\pm\frac{\sqrt{(r^2+a^2)^2-a^2\Delta}}{\Delta}.
\end{equation}
A convenient representative of the corresponding conformal class is\footnote{The geometric origin of this reduction can also be understood directly from the Kerr metric. In the prolate asymptotic sector under consideration, the angular eigenfunctions concentrate near the equatorial plane. Moreover, at fixed $m$, the ratio $m/\omega$ tends to zero, so the leading rays have vanishing azimuthal angular momentum relative to their energy.}
\begin{equation}\label{eq:metric-2D2}
\dd s^2_{\text{pr}}=-\frac{\Delta}{\sqrt{(r^2+a^2)^2-a^2\Delta}}\,\dd t^2+\frac{\sqrt{(r^2+a^2)^2-a^2\Delta}}{\Delta}\,\dd r^2.
\end{equation}
The radial tortoise coordinate adapted to this metric is defined by
\begin{equation}\label{eq:2DKerr-tortoise}
\frac{\dd x}{\dd r}=\frac{\sqrt{(r^2+a^2)^2-a^2\Delta}}{\Delta}.
\end{equation}
In particular, the prolate and oblate geometries have the same horizon logarithms but different global optical coordinates.

The null curves \eqref{eq:radial-characteristics} govern the propagation of singularities in the 2D optical metric. Solving  \eqref{eq:radial-characteristics} gives geodesics that reach the complex turning points at the roots of the radial polynomial under the square root. These are bouncing geodesics in the same sense as those found in Schwarzschild \cite{Arnaudo:2026tcy}, albeit complex. Precisely, the turning point is given by $r_c$ as given in \eqref{eq:intro-rc}. The geodesic after the bounce therefore corresponds to a complex $t$ singularity of the Green's function at fixed $\ell, m, r, r', t'$. It is this singularity that is responsible for setting the radius of convergence of the prolate QNM sum.

\begin{figure}[h!]
\centering
\resizebox{0.98\linewidth}{!}{%
\begin{tikzpicture}[x=1.55cm,y=1cm,every node/.style={font=\footnotesize},>=Stealth]
\draw[->,gray!75] (-.3,0)--(4.05,0);
\node[gray!75,below] at (3.9,-.08) {$\mathrm{Re}\,r$};
\draw[->,gray!75] (0,-1.85)--(0,1.95);
\node[gray!75,above] at (0,2.0) {$\mathrm{Im}\,r$};

\draw[orange!80!black,line width=1.6pt] (2.2,0)--(3.65,0);
\node[orange!80!black,below] at (3.15,-.22) {exterior};
\fill[black] (.45,0) circle (1.25pt);
\fill[black] (2.2,0) circle (1.25pt);
\node[below=4pt] at (.45,0) {$r_-$};
\node[below=4pt] at (2.2,0) {$r_+$};

\draw[blue!75!black,line width=1.05pt,
  postaction={decorate},
  decoration={markings,mark=at position .53 with {\arrow{Stealth}}}]
  (1.05,-1.5) .. controls (2.5,-1.5) and (3.05,-.8) .. (3.05,0)
  .. controls (3.05,.8) and (2.5,1.5) .. (1.05,1.5);
\node[blue!75!black,anchor=west] at (3.13,.95) {$\gamma_{\text{pr}}$};

\fill[red!75!black] (1.05,1.5) circle (1.65pt);
\fill[red!75!black] (1.05,-1.5) circle (1.65pt);
\node[above left,red!75!black] at (1.05,1.5) {$r_c$};
\node[below left,red!75!black] at (1.05,-1.5) {$\bar r_c$};
\end{tikzpicture}%
}
\captionsetup{width=\linewidth,justification=justified,singlelinecheck=false}
\caption{The contour $\gamma_{\text{pr}}$ in the complex $r$-plane, oriented from $\bar r_c$ to $r_c$ and crossing the real axis outside the outer horizon, at $r>r_+$ (shown in orange). Red points mark the conjugate complex turning points and black points the horizons.}
\label{fig:complex-separation}
\end{figure}
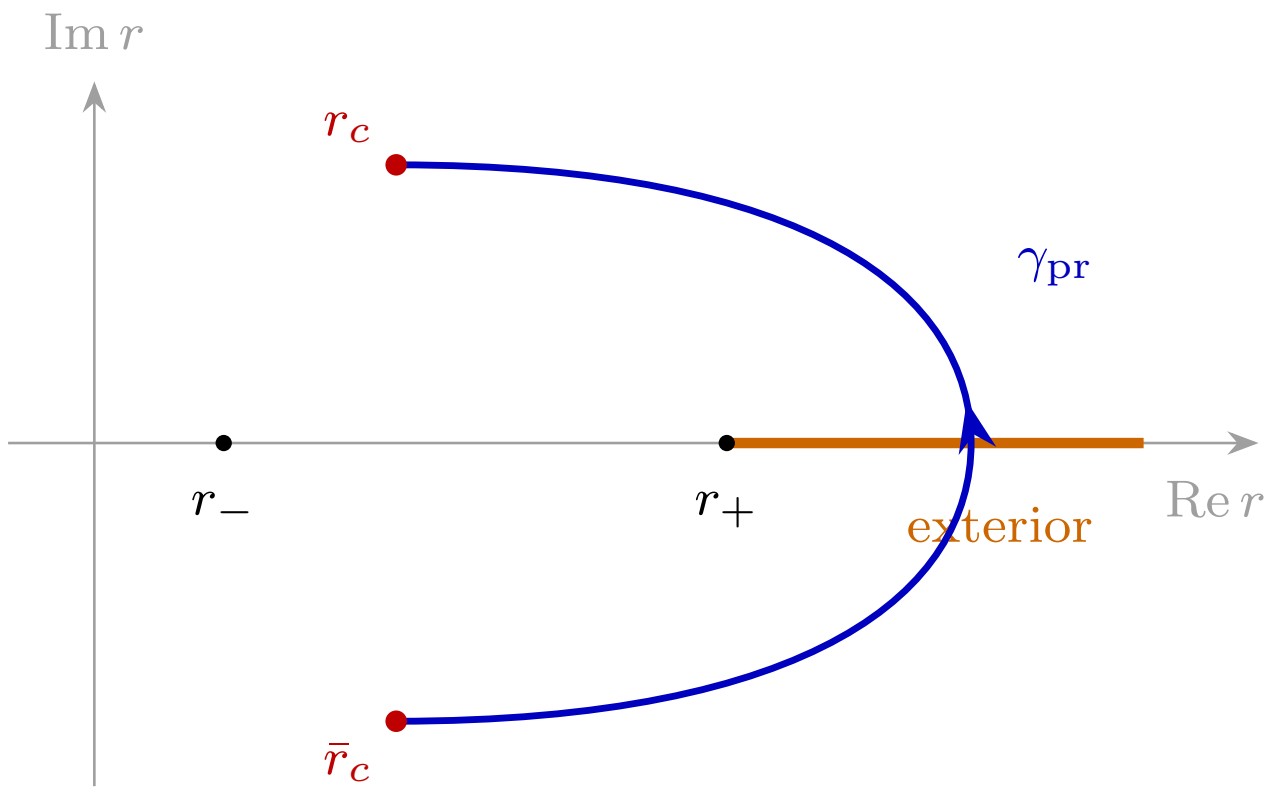

For an insertion at $t', r'$ and where $r,r'>r^{\text{bounce}}_{\text{prolate}}$, the relevant geodesic is an ingoing one, which before the turning point is $t-t' = -(x-x')$ where $x' = x(r')$. After the turning point, the geodesic reads
\be
t - t' = x + x' - 2x(r_c). \label{prolatebouncingsingularity}
\ee
which corresponds to complex singularities as $r_c$ is complex. 
Additional candidate singularities related to \eqref{prolatebouncingsingularity} arise by allowing additional windings around the various cycles. One cycle is associated with the complex-conjugate turning points $r_c$ and $\bar r_c$. This cycle follows the contour $\gamma_{\text{pr}}$ connecting the two turning points through the real exterior and returns on the other sheet of the square root, see Figure \ref{fig:complex-separation}. Choosing conjugate continuations from the exterior, its optical period is
\be
2\bigl[x(r_c)-x(\bar r_c)\bigr]=4\ii\,\mathrm{Im}x(r_c).
\ee
Other cycles come from log branch points at the horizons $r_\pm$, which contribute integer multiples of the periods
\be
\frac{2\pi i}{\kappa_\pm}.
\ee
The corresponding candidate image singularities therefore occur at
\be
t-t'=x+x'-2x(r_c)+4k\ii\,\mathrm{Im}x(r_c) +\sum_{\pm} k_\pm\frac{2\pi i}{\kappa_\pm},\quad k,k_\pm\in\mathbb Z. \label{prolatesingsk}
\ee
Since these additional period contributions are purely imaginary, they have the same modulus in the exponential time variable of the prolate QNM expansion and do not alter its radius of convergence. These are the analogues of the future-bouncing singularities in Schwarzschild.

Similarly, when $r,r'<r^{\text{bounce}}$ the relevant geodesic is an outgoing one, $t-t' = x-x'$ before the turning point, and after the turning point gives the past-bouncing singularities
\be
t-t'=-x-x'+2x(r_c)+4k\ii\,\mathrm{Im}x(r_c)+ \sum_{\pm}k_\pm\frac{2\pi i}{\kappa_\pm},\quad k,k_\pm\in\mathbb Z. \label{prolatesingskpast}
\ee

The singularity analysis presented thus far predicts the location of $r^\text{bounce}_\text{prolate}$, which we also verify shortly using QNM asymptotics. In particular with the prolate family of QNMs behaving asymptotically as \cite{Keshet:2007nv}
\begin{equation}\label{eq:qnm-prolate}
\omega_{n+1}^{\mathrm{pr}}-\omega_{n}^{\mathrm{pr}}\sim-\ii\rho,
\end{equation}
\begin{equation}\label{eq:quoted-prolate-spacing}
\rho=\frac{\pi}{\displaystyle\mathrm{Im}\int_{\gamma_{\text{pr}}}
\frac{\sqrt{(r^2+a^2)^2-a^2\Delta}}{\Delta}\,\dd r},
\end{equation}
following \cite{Arnaudo:2026tcy}, the asymptotic QNM sum becomes a power series in the variable $X^{\text{prolate}} = e^{-\rho t}$ up to phase factors, and therefore in cases where the future-bounced singularity is closest to $X^\text{prolate}=0$ the convergence radius in $X^{\text{prolate}}$ is 
\be
|X^{\text{prolate}}| = e^{-\rho(t' + x + x' - 2\mathrm{Re}(x(r_c)))}
\ee
as determined by the singularity \eqref{prolatebouncingsingularity}.
Hence in the physical domain the convergence is restricted to the region $e^{-\rho t} < e^{-\rho(t' + x + x' - 2\mathrm{Re}(x(r_c)))}$. In other words, to the future of the right-moving null ray
\be
t-t' = x+x' - 2\mathrm{Re}(x(r_c)).
\ee
This ray intersects the ingoing ray at 
\be
x(r^\text{bounce}_\text{prolate}) = \mathrm{Re}(x(r_c))
\ee
thus defining the bounce radius $r^\text{bounce}_\text{prolate}$. Similar arguments apply when the past-bounced singularity is the closest to $X^\text{prolate}=0$.

Let us now turn to the asymptotic QNM sum in order to verify its convergence and singularities are consistent with the above bouncing geodesic computation. 

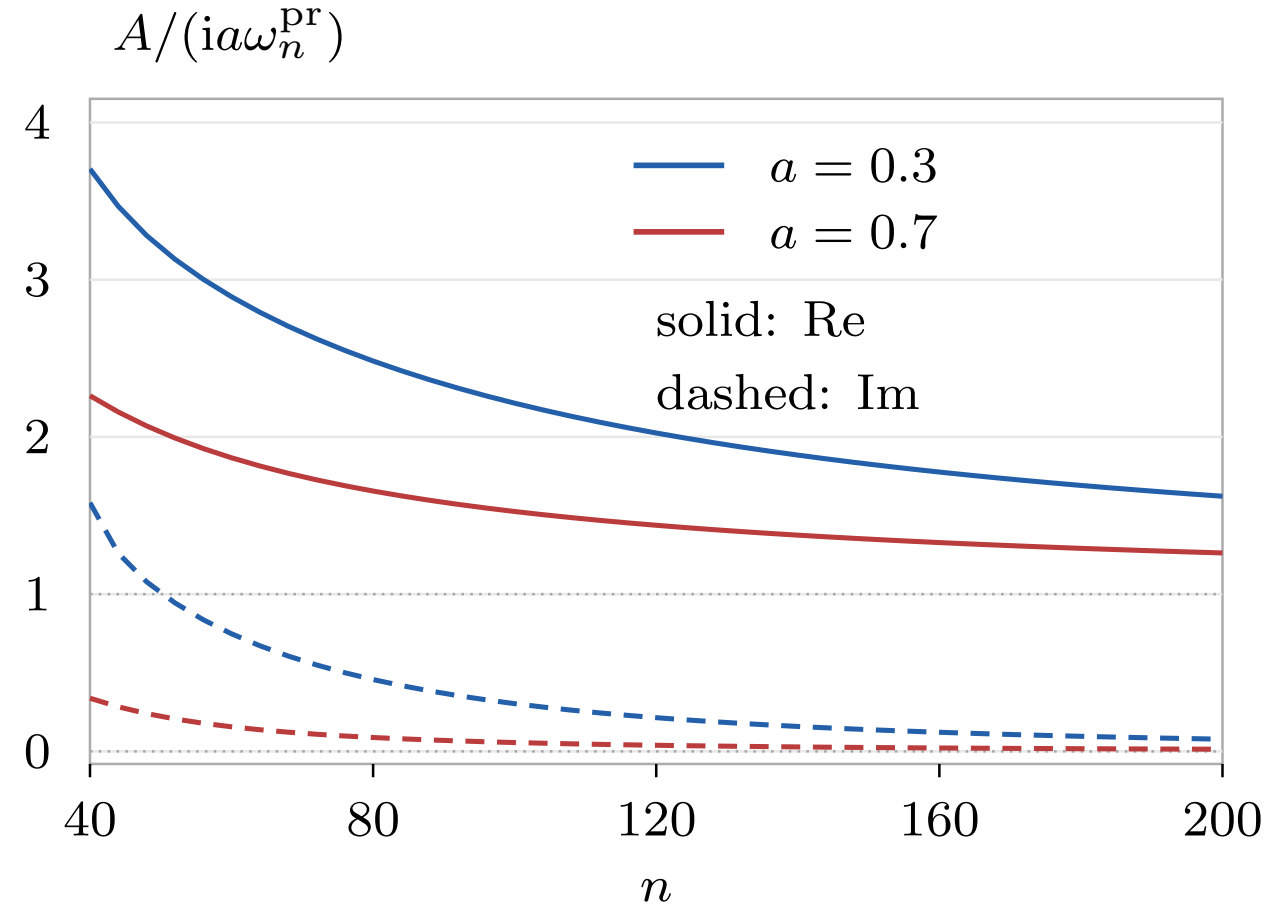
\begin{figure}[h!]
\centering
\resizebox{0.98\linewidth}{!}{%
\begin{tikzpicture}[x=6.3cm,y=3.7cm,every node/.style={font=\footnotesize}]
\definecolor{angblue}{RGB}{36,96,170}
\definecolor{angred}{RGB}{187,60,60}
\draw[gray!65] (0,0) rectangle (1,1);
\draw[gray!18] (0,0.0189125)--(1,0.0189125);
\node[left] at (-.015,0.0189125) {$ 0 $};
\draw[gray!18] (0,0.2553191)--(1,0.2553191);
\node[left] at (-.015,0.2553191) {$ 1 $};
\draw[gray!18] (0,0.4917258)--(1,0.4917258);
\node[left] at (-.015,0.4917258) {$ 2 $};
\draw[gray!18] (0,0.7281324)--(1,0.7281324);
\node[left] at (-.015,0.7281324) {$ 3 $};
\draw[gray!18] (0,0.9645390)--(1,0.9645390);
\node[left] at (-.015,0.9645390) {$ 4 $};
\draw (0.0000000,0)--(0.0000000,-.02);
\node[below] at (0.0000000,-.025) {$ 40 $};
\draw (0.2500000,0)--(0.2500000,-.02);
\node[below] at (0.2500000,-.025) {$ 80 $};
\draw (0.5000000,0)--(0.5000000,-.02);
\node[below] at (0.5000000,-.025) {$ 120 $};
\draw (0.7500000,0)--(0.7500000,-.02);
\node[below] at (0.7500000,-.025) {$ 160 $};
\draw (1.0000000,0)--(1.0000000,-.02);
\node[below] at (1.0000000,-.025) {$ 200 $};
\draw[gray!70,densely dotted] (0,0.0189125)--(1,0.0189125);
\draw[gray!70,densely dotted] (0,0.2553191)--(1,0.2553191);
\begin{scope}
\clip (0,0) rectangle (1,1);
\draw[angblue,line width=.8pt] plot coordinates {
(0.0000000,0.8945910) (0.0250000,0.8384411) (0.0500000,0.7945393) (0.0750000,0.7589045)
(0.1000000,0.7287174) (0.1250000,0.7024279) (0.1500000,0.6791005) (0.1750000,0.6581331)
(0.2000000,0.6391160) (0.2250000,0.6217537) (0.2500000,0.6058228) (0.2750000,0.5911465)
(0.3000000,0.5775807) (0.3250000,0.5650047) (0.3500000,0.5533161) (0.3750000,0.5424265)
(0.4000000,0.5322592) (0.4250000,0.5227472) (0.4500000,0.5138312) (0.4750000,0.5054591)
(0.5000000,0.4975844) (0.5250000,0.4901658) (0.5500000,0.4831662) (0.5750000,0.4765525)
(0.6000000,0.4702947) (0.6250000,0.4643659) (0.6500000,0.4587419) (0.6750000,0.4534004)
(0.7000000,0.4483214) (0.7250000,0.4434868) (0.7500000,0.4388798) (0.7750000,0.4344853)
(0.8000000,0.4302893) (0.8250000,0.4262791) (0.8500000,0.4224430) (0.8750000,0.4187701)
(0.9000000,0.4152507) (0.9250000,0.4118754) (0.9500000,0.4086359) (0.9750000,0.4055244)
(1.0000000,0.4025335)
};
\draw[angblue,line width=.8pt,densely dashed] plot coordinates {
(0.0000000,0.3927269) (0.0250000,0.3167312) (0.0500000,0.2738547) (0.0750000,0.2421995)
(0.1000000,0.2168251) (0.1250000,0.1956949) (0.1500000,0.1777351) (0.1750000,0.1622821)
(0.2000000,0.1488762) (0.2250000,0.1371736) (0.2500000,0.1269045) (0.2750000,0.1178516)
(0.3000000,0.1098369) (0.3250000,0.1027129) (0.3500000,0.0963571) (0.3750000,0.0906664)
(0.4000000,0.0855543) (0.4250000,0.0809474) (0.4500000,0.0767833) (0.4750000,0.0730088)
(0.5000000,0.0695782) (0.5250000,0.0664521) (0.5500000,0.0635967) (0.5750000,0.0609824)
(0.6000000,0.0585835) (0.6250000,0.0563778) (0.6500000,0.0543455) (0.6750000,0.0524693)
(0.7000000,0.0507342) (0.7250000,0.0491266) (0.7500000,0.0476346) (0.7750000,0.0462477)
(0.8000000,0.0449565) (0.8250000,0.0437525) (0.8500000,0.0426283) (0.8750000,0.0415770)
(0.9000000,0.0405926) (0.9250000,0.0396697) (0.9500000,0.0388034) (0.9750000,0.0379892)
(1.0000000,0.0372231)
};
\draw[angred,line width=.8pt] plot coordinates {
(0.0000000,0.5536438) (0.0250000,0.5291716) (0.0500000,0.5082029) (0.0750000,0.4900614)
(0.1000000,0.4742314) (0.1250000,0.4603129) (0.1500000,0.4479909) (0.1750000,0.4370143)
(0.2000000,0.4271808) (0.2250000,0.4183255) (0.2500000,0.4103133) (0.2750000,0.4030322)
(0.3000000,0.3963887) (0.3250000,0.3903044) (0.3500000,0.3847129) (0.3750000,0.3795578)
(0.4000000,0.3747908) (0.4250000,0.3703703) (0.4500000,0.3662605) (0.4750000,0.3624301)
(0.5000000,0.3588520) (0.5250000,0.3555024) (0.5500000,0.3523603) (0.5750000,0.3494072)
(0.6000000,0.3466268) (0.6250000,0.3440044) (0.6500000,0.3415271) (0.6750000,0.3391833)
(0.7000000,0.3369625) (0.7250000,0.3348555) (0.7500000,0.3328537) (0.7750000,0.3309494)
(0.8000000,0.3291359) (0.8250000,0.3274068) (0.8500000,0.3257563) (0.8750000,0.3241794)
(0.9000000,0.3226711) (0.9250000,0.3212271) (0.9500000,0.3198434) (0.9750000,0.3185164)
(1.0000000,0.3172426)
};
\draw[angred,line width=.8pt,densely dashed] plot coordinates {
(0.0000000,0.0988134) (0.0250000,0.0858987) (0.0500000,0.0757734) (0.0750000,0.0677084)
(0.1000000,0.0611937) (0.1250000,0.0558654) (0.1500000,0.0514584) (0.1750000,0.0477763)
(0.2000000,0.0446718) (0.2250000,0.0420323) (0.2500000,0.0397710) (0.2750000,0.0378202)
(0.3000000,0.0361264) (0.3250000,0.0346470) (0.3500000,0.0333478) (0.3750000,0.0322011)
(0.4000000,0.0311842) (0.4250000,0.0302783) (0.4500000,0.0294682) (0.4750000,0.0287408)
(0.5000000,0.0280855) (0.5250000,0.0274930) (0.5500000,0.0269556) (0.5750000,0.0264669)
(0.6000000,0.0260210) (0.6250000,0.0256132) (0.6500000,0.0252393) (0.6750000,0.0248957)
(0.7000000,0.0245791) (0.7250000,0.0242869) (0.7500000,0.0240166) (0.7750000,0.0237661)
(0.8000000,0.0235335) (0.8250000,0.0233172) (0.8500000,0.0231157) (0.8750000,0.0229276)
(0.9000000,0.0227518) (0.9250000,0.0225873) (0.9500000,0.0224330) (0.9750000,0.0222883)
(1.0000000,0.0221523)
};
\end{scope}
\node[anchor=west] at (0,1.095) {$ A/(\ii a\omega_n^{\text{pr}}) $};
\node[below] at (.5,-.14) {$n$};
\draw[angblue,line width=.9pt] (.48,.90)--(.56,.90);
\node[anchor=west] at (.58,.90) {$a=0.3$};
\draw[angred,line width=.9pt] (.48,.80)--(.56,.80);
\node[anchor=west] at (.58,.80) {$a=0.7$};
\node[anchor=west] at (.48,.67) {solid: $\mathrm{Re}$};
\node[anchor=west] at (.48,.56) {dashed: $\mathrm{Im}$};
\end{tikzpicture}%
}
\caption{Angular eigenvalues on the prolate QNM sequence. Solid and dashed curves show the real and imaginary parts of $A/(\ii a\omega_n^{\text{pr}})$, respectively; blue is $a=0.3$ and red is $a=0.7$. }
\label{fig:prolate-angular-data}
\end{figure}

We numerically verified that the optical metric \eqref{eq:metric-2D2} is selected by the prolate sequence, that is, that the angular eigenvalues have the corresponding asymptotic behaviour when evaluated along this branch of modes at large overtone number. At selected rotations, $a=0.3,0.7$, we have computed 201 prolate modes, labelled by $n=0,\ldots,200$. 
This sequence has the asymptotic spacing \eqref{eq:qnm-prolate}, while its real part approaches the negative limiting value predicted in \cite{Keshet:2007nv}:
\begin{equation}
\lim_{n\to\infty}\mathrm{Re}\,\omega_n^{\mathrm{pr}}=\mathrm{Re}\,\omega_0\simeq
\begin{cases}
-0.125425,&a=0.3,\\
-0.114982,&a=0.7.
\end{cases}
\end{equation}
The angular asymptotics give $A(a\omega_n^{\mathrm{pr}})\sim\ii a\omega_n^{\mathrm{pr}}$ \cite{Keshet:2012uq}. 
Figure \ref{fig:prolate-angular-data} shows the first ratio $A/(\mathrm{i}a\omega_n^{\text{pr}})$ evaluated at the computed frequencies. The approach is slower at $a=0.3$, where $|a\omega_n|$ is smaller at the same overtone number.

The positive spacing predicted by \eqref{eq:quoted-prolate-spacing} is $\rho=0.252920$ at $a=0.3$ and $\rho=0.255364$ at $a=0.7$. At the last pair of computed modes, $n=199,200$, the imaginary parts of the frequency are spaced by $0.2528156$ and $0.2553380$, respectively in good agreement. The spacings of the real parts have magnitudes below $1.4\times10^{-5}$. Thus these values approach the asymptotic values $\omega_n^{\mathrm{pr}}\sim\omega_0-\ii\rho n$.

To verify the region of convergence of the QNM sum we must also study the residues. 
We consider the QNM residues at fixed exterior radii $r>r'$, with both points beyond the location $x(r')>\mathrm{Re}\,x(r_c)$. Using the up-up representation in \eqref{eq:three-residues}, the numerical data are compatible with the leading behaviour
\begin{widetext}
\begin{equation}\label{eq:prolate-sum}
\begin{aligned}
G_{\mathrm{pr}}^{\mathrm{leading}}(t-t';r,r')&\propto e^{-\ii\omega_0(t-t')}\sum_{n=1}^{\infty}\frac{(-1)^n}{n}\exp\left[-\rho n\left(t-t'-x-x'+2\mathrm{Re}\,x(r_c)\right)\right]\\
&=-e^{-\ii\omega_0(t-t')}\log\left(1+\exp\left[-\rho\left(t-t'-x-x'+2\mathrm{Re}\,x(r_c)\right)\right]\right),
\end{aligned}
\end{equation}
\end{widetext}
which shows logarithmic branch point singularities at
\begin{equation}\label{eq:latest-pr-images}
t-t'=x+x'-2\mathrm{Re}\,x( r_c)+\frac{(2k+1)\pi\ii}{\rho},\qquad k\in\mathbb Z,
\end{equation}
in agreement with a subset of the candidate singularity locations given by the bouncing geodesic analysis in \eqref{prolatesingsk} since
\be
x( r_c)=\mathrm{Re}\,x( r_c)+\mathrm{i}\,\frac{\pi}{2\rho}.
\ee
The singularities form a discrete sequence on a vertical line in the complex $t-t'$ plane and specify the corresponding open half-plane of absolute convergence as
\begin{equation}\label{eq:pr-outer-convergence}
\mathrm{Re}\left(t-t'\right)>x+x'-2\mathrm{Re}\,x(r_c).
\end{equation}
The residue asymptotics thus support the effective exterior bounce radius defined by $x(r^{\mathrm{bounce}}_\text{prolate})=\mathrm{Re}\,x(r_c)$ matching the bouncing geodesic derivation. 

When $r_+<r,r' < r^\text{bounce}_\text{prolate}$, the in-in form of \eqref{eq:three-residues} is convenient. The radial data support the reversed real optical distance with the same alternating $1/n$ prefactor:
\begin{equation}\label{eq:inner-prolate-residue}
\begin{aligned}
\underset{\omega=\omega_n^{\mathrm{pr}}}{\mathrm{Res}}\widetilde G
&=\frac{C_{\mathrm{up},\mathrm{in}}(\omega_n^{\mathrm{pr}})R_{\mathrm{in},n}(r)R_{\mathrm{in},n}(r')}{\left.\frac{\dd W}{\dd\omega}\right|_{\omega_n^{\mathrm{pr}}}}\\
&\sim\text{constant}\,\frac{(-1)^n}{n}e^{\rho n\left(2\mathrm{Re}\,x(r_c)-x-x'\right)}.
\end{aligned}
\end{equation}
The corresponding sum at leading order gives
\begin{equation}
-e^{-\ii\omega_0(t-t')}\log\left(1+e^{-\rho\left(t-t'-2\mathrm{Re}\,x(r_c)+x+x'\right)}\right),
\end{equation}
with logarithmic branch points at
\begin{equation}\label{eq:inner-prolate-bounces}
t-t'=2\mathrm{Re}\,x(r_c)-x-x'+\frac{(2k+1)\pi\ii}{\rho}.
\end{equation}
which coincide with a subset of the candidate past-bouncing singularities given by the geodesic analysis \eqref{prolatesingskpast}.
The open half-plane of convergence is now $\mathrm{Re}\left(t-t'\right)>2\mathrm{Re}\,x(r_c)-x-x'$.

For points on opposite sides of the bounce, the up-in representation depends on the direct optical distance $x-x'$. The mixed data are compatible with a $1/n$ coefficient without alternation; the corresponding logarithm is proportional to $-e^{-\ii\omega_0(t-t')}\log[1-e^{-\rho(t-t'-x+x')}]$. This has singularities corresponding to the direct lightray at
$t-t'=x-x'+2k\pi\ii/\rho$, $k\in\mathbb{Z}$.
Overall, combining the three configurations, we find the following region of convergence:
\begin{equation}\label{eq:pr-family-region}
\mathrm{Re}\left(t-t'\right)>|x-\mathrm{Re}\,x(r_c)|+|x'-\mathrm{Re}\,x(r_c)|.
\end{equation}

\section{Oblate QNM family} \label{sec:oblate}
The second case to be analysed is the \emph{oblate} asymptotic sector, where the angular eigenvalues grow as $A\sim-a^2\omega^2$. In particular,
\begin{equation}\label{eq:th-angular-check}
\frac{\lambda(a\omega_n^{\text{ob}})}{(\omega_n^{\text{ob}})^2}\to 0.
\end{equation} 
Due to the cancellation of the quadratic term in $\lambda$, the potential in \eqref{eq:normalform} reduces to
\begin{equation}
V\sim\omega^2\frac{\left(r^2+a^2\right)^2}{\Delta^2}.
\end{equation}
In this case, the radial characteristics are 
\begin{equation}\label{eq:oblate-characteristics}
\frac{\dd t}{\dd r}=\pm\frac{r^2+a^2}{\Delta}
\end{equation}
and are represented by the metric obtained by restricting \eqref{BL} to the symmetry axis,
\begin{equation}\label{eq:oblate-metric}
\dd s^2_{\mathrm{ob}}=-\frac{\Delta}{r^2+a^2}\dd t^2+\frac{r^2+a^2}{\Delta}\dd r^2.
\end{equation}
Its optical coordinate is the ordinary tortoise coordinate 
\begin{equation}
\frac{\dd r_*}{\dd r}=\frac{r^2+a^2}{\Delta}.
\end{equation}

The null curves \eqref{eq:oblate-characteristics} govern the propagation of singularities\footnote{These curves can be realised as limits of spacelike geodesics connecting regions $\mathrm I$ and $\mathrm I_L$. When the energy of these geodesics becomes large, the turning point becomes complex.} in the oblate metric \eqref{eq:oblate-metric}.
The complex turning points of the null curves are located at $r = \pm\ii a$.

For an insertion at $t',r'$ with $r,r'>r^{\text{bounce}}_\text{oblate}$, the ingoing null line before the turning point is $t-t'=-(r_*-r'_*)$, where $r'_*=r_*(r')$. After the turning point, the null line reads
\begin{equation}
t-t'=r_*+r'_*-2r_*(\ii a),
\end{equation}
which corresponds to complex singularities in $t$.
Unlike the prolate case, $\pm\ii a$ are not branch points of the characteristics \eqref{eq:oblate-characteristics}, since $(r^2+a^2)/\Delta$ is single valued, and $r_*(\ii a)-r_*(-\ii a)$ is not a period. The only periods of $r_*$ come from the horizon logarithms: each winding around $r=r_\pm$ shifts $r_*$ by $\ii\pi/\kappa_\pm$. The corresponding candidate image singularities are
\begin{equation}
t-t'=r_*+r_*'-2r_*(\ii a)+\sum_\pm \frac{2\pi k_\pm\ii}{\kappa_\pm},\qquad k_\pm\in\mathbb Z. \label{oblatesingsk}
\end{equation}
As in the prolate case, the periods are purely imaginary and will not alter the radius of convergence of the QNM sum.

As in the prolate case, when $r,r'<r^{\text{bounce}}_\text{oblate}$ the relevant geodesic is an outgoing one, and after the turning point gives rise to candidate past-bouncing singularities at
\be
t-t'=-r_*-r_*'+2r_*(\ii a)+\sum_{\pm}\frac{2\pi k_\pm\ii}{\kappa_\pm},\qquad k_\pm\in\mathbb Z. \label{oblatesingskpast}
\ee

From this singularity information we can now obtain $r^\text{bounce}_\text{oblate}$. The oblate family of QNMs satisfies the asymptotic behaviour \cite{Keshet:2012uq}
\begin{equation}\label{eq:qnm-oblate}
\omega_{n+1}^{\text{ob}}-\omega_n^{\text{ob}}\to-\ii\kappa_+,\qquad
\mathrm{Re}\,\omega_n^{\text{ob}}\to m\Omega_+.
\end{equation}
Hence, the asymptotic QNM sum becomes a power series in the variable $X^{\text{oblate}}=e^{-\kappa_+ t}$ at fixed $t', r, r'$ up to phases, and the convergence radius in the case where the future-bouncing singularity is the closest to  $X^{\text{oblate}} = 0$ is
\be
|X^{\text{oblate}}|=e^{-\kappa_+\left(t'+r_*+r'_*-2\mathrm{Re}\,r_*(\ii a)\right)}.
\ee
As a consequence, the domain of convergence is restricted to the future of the right-moving null ray
\be
t-t'=r_*+r'_*-2\mathrm{Re}\,r_*(\ii a),
\ee
which intersects the ingoing ray at
\be
r_*(r^{\text{bounce}}_{\text{oblate}})=\mathrm{Re}\,r_*(\ii a),
\ee
that thus defines the bounce radius $r^{\text{bounce}}_{\text{oblate}}$. Similar arguments apply when the past-bounced singularity is the closest to $X^\text{oblate}=0$.

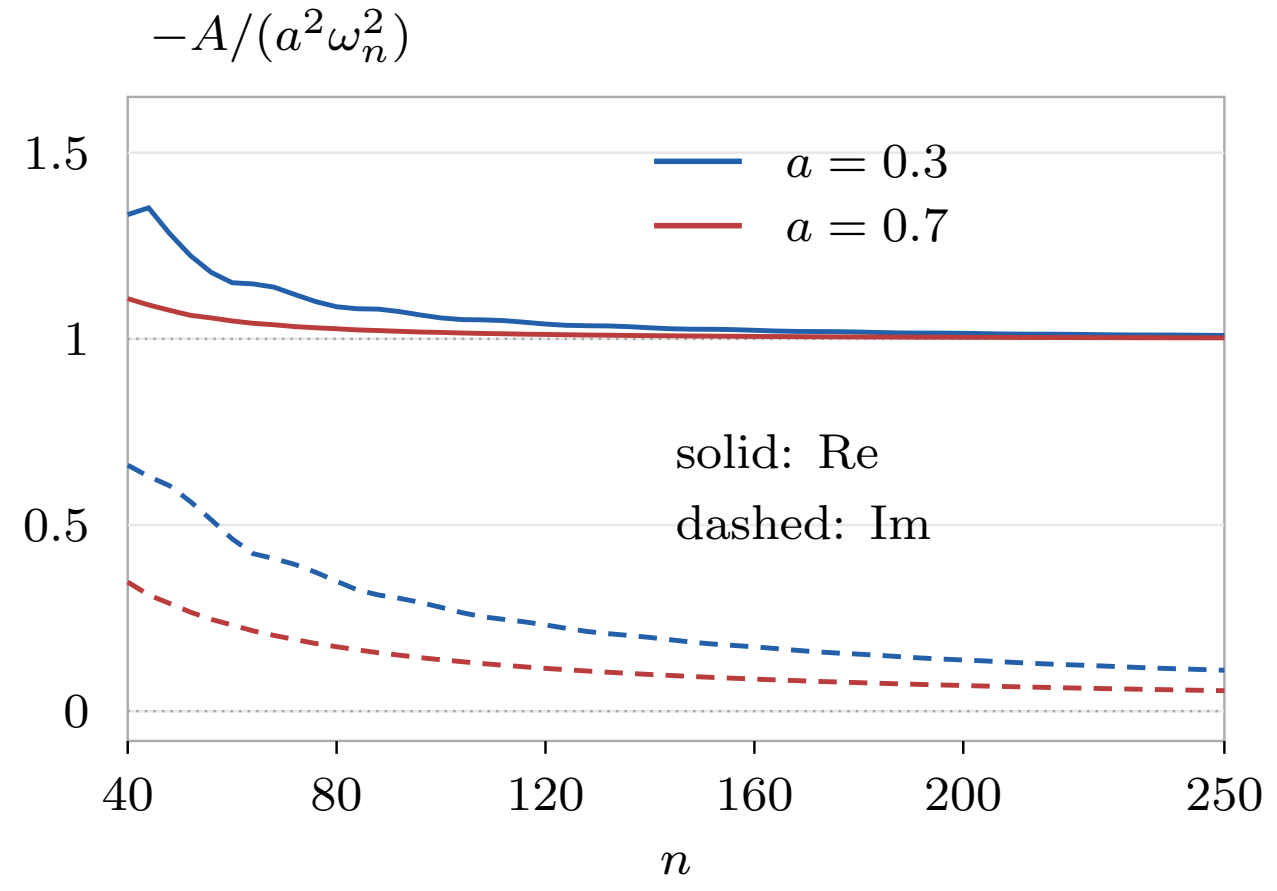
\begin{figure}[h!]
\centering
\resizebox{0.98\linewidth}{!}{%
\begin{tikzpicture}[x=6.3cm,y=3.7cm,every node/.style={font=\footnotesize}]
\definecolor{angblue}{RGB}{36,96,170}
\definecolor{angred}{RGB}{187,60,60}
\draw[gray!65] (0,0) rectangle (1,1);
\draw[gray!18] (0,0.0462428)--(1,0.0462428);
\node[left] at (-.015,0.0462428) {$ 0 $};
\draw[gray!18] (0,0.3352601)--(1,0.3352601);
\node[left] at (-.015,0.3352601) {$ 0.5 $};
\draw[gray!18] (0,0.6242775)--(1,0.6242775);
\node[left] at (-.015,0.6242775) {$ 1 $};
\draw[gray!18] (0,0.9132948)--(1,0.9132948);
\node[left] at (-.015,0.9132948) {$ 1.5 $};
\draw (0.0000000,0)--(0.0000000,-.02);
\node[below] at (0.0000000,-.025) {$ 40 $};
\draw (0.1904762,0)--(0.1904762,-.02);
\node[below] at (0.1904762,-.025) {$ 80 $};
\draw (0.3809524,0)--(0.3809524,-.02);
\node[below] at (0.3809524,-.025) {$ 120 $};
\draw (0.5714286,0)--(0.5714286,-.02);
\node[below] at (0.5714286,-.025) {$ 160 $};
\draw (0.7619048,0)--(0.7619048,-.02);
\node[below] at (0.7619048,-.025) {$ 200 $};
\draw (1.0000000,0)--(1.0000000,-.02);
\node[below] at (1.0000000,-.025) {$ 250 $};
\draw[gray!70,densely dotted] (0,0.0462428)--(1,0.0462428);
\draw[gray!70,densely dotted] (0,0.6242775)--(1,0.6242775);
\begin{scope}
\clip (0,0) rectangle (1,1);
\draw[angblue,line width=.8pt] plot coordinates {
(0.0000000,0.8172120) (0.0190476,0.8277656) (0.0380952,0.7879919) (0.0571429,0.7535717)
(0.0761905,0.7274587) (0.0952381,0.7114427) (0.1142857,0.7096076) (0.1333333,0.7046170)
(0.1523810,0.6931791) (0.1714286,0.6821553) (0.1904762,0.6741286) (0.2095238,0.6710287)
(0.2285714,0.6704684) (0.2476190,0.6666862) (0.2666667,0.6614013) (0.2857143,0.6567926)
(0.3047619,0.6542420) (0.3238095,0.6537911) (0.3428571,0.6526705) (0.3619048,0.6500855)
(0.3809524,0.6472826) (0.4000000,0.6453300) (0.4190476,0.6447062) (0.4380952,0.6443912)
(0.4571429,0.6431839) (0.4761905,0.6414612) (0.4952381,0.6399959) (0.5142857,0.6393064)
(0.5333333,0.6391532) (0.5523810,0.6386396) (0.5714286,0.6376062) (0.5904762,0.6365271)
(0.6095238,0.6358602) (0.6285714,0.6356823) (0.6476190,0.6354852) (0.6666667,0.6348987)
(0.6857143,0.6341261) (0.7047619,0.6335296) (0.7238095,0.6332989) (0.7428571,0.6332161)
(0.7619048,0.6329103) (0.7809524,0.6323796) (0.8000000,0.6318754) (0.8190476,0.6316103)
(0.8380952,0.6315446) (0.8571429,0.6314014) (0.8761905,0.6310573) (0.8952381,0.6306519)
(0.9142857,0.6303785) (0.9333333,0.6302925) (0.9523810,0.6302299) (0.9714286,0.6300233)
(0.9904762,0.6297140) (1.0000000,0.6295688)
};
\draw[angblue,line width=.8pt,densely dashed] plot coordinates {
(0.0000000,0.4282298) (0.0190476,0.4107591) (0.0380952,0.3962992) (0.0571429,0.3716515)
(0.0761905,0.3433984) (0.0952381,0.3137884) (0.1142857,0.2907846) (0.1333333,0.2831872)
(0.1523810,0.2740690) (0.1714286,0.2618797) (0.1904762,0.2478747) (0.2095238,0.2344997)
(0.2285714,0.2266507) (0.2476190,0.2215081) (0.2666667,0.2151451) (0.2857143,0.2073779)
(0.3047619,0.1991782) (0.3238095,0.1927735) (0.3428571,0.1888049) (0.3619048,0.1849271)
(0.3809524,0.1801829) (0.4000000,0.1748412) (0.4190476,0.1699623) (0.4380952,0.1665721)
(0.4571429,0.1638361) (0.4761905,0.1607137) (0.4952381,0.1570713) (0.5142857,0.1534022)
(0.5333333,0.1504961) (0.5523810,0.1483198) (0.5714286,0.1461132) (0.5904762,0.1435363)
(0.6095238,0.1407668) (0.6285714,0.1383173) (0.6476190,0.1364553) (0.6666667,0.1347775)
(0.6857143,0.1328900) (0.7047619,0.1307807) (0.7238095,0.1287474) (0.7428571,0.1271066)
(0.7619048,0.1257385) (0.7809524,0.1243032) (0.8000000,0.1226781) (0.8190476,0.1210070)
(0.8380952,0.1195563) (0.8571429,0.1183769) (0.8761905,0.1172381) (0.8952381,0.1159673)
(0.9142857,0.1146011) (0.9333333,0.1133286) (0.9523810,0.1122764) (0.9714286,0.1113306)
(0.9904762,0.1103173) (1.0000000,0.1097679)
};
\draw[angred,line width=.8pt] plot coordinates {
(0.0000000,0.6867818) (0.0190476,0.6769868) (0.0380952,0.6686690) (0.0571429,0.6609031)
(0.0761905,0.6567795) (0.0952381,0.6521369) (0.1142857,0.6484706) (0.1333333,0.6461959)
(0.1523810,0.6433692) (0.1714286,0.6415143) (0.1904762,0.6399711) (0.2095238,0.6382139)
(0.2285714,0.6371819) (0.2476190,0.6360242) (0.2666667,0.6349441) (0.2857143,0.6342721)
(0.3047619,0.6333901) (0.3238095,0.6327339) (0.3428571,0.6322158) (0.3619048,0.6315640)
(0.3809524,0.6311537) (0.4000000,0.6307143) (0.4190476,0.6302555) (0.4380952,0.6299705)
(0.4571429,0.6295947) (0.4761905,0.6292857) (0.4952381,0.6290551) (0.5142857,0.6287471)
(0.5333333,0.6285404) (0.5523810,0.6283327) (0.5714286,0.6280959) (0.5904762,0.6279480)
(0.6095238,0.6277573) (0.6285714,0.6275860) (0.6476190,0.6274645) (0.6666667,0.6272967)
(0.6857143,0.6271765) (0.7047619,0.6270640) (0.7238095,0.6269262) (0.7428571,0.6268387)
(0.7619048,0.6267306) (0.7809524,0.6266251) (0.8000000,0.6265534) (0.8190476,0.6264530)
(0.8380952,0.6263761) (0.8571429,0.6263092) (0.8761905,0.6262222) (0.8952381,0.6261655)
(0.9142857,0.6260992) (0.9333333,0.6260295) (0.9523810,0.6259834) (0.9714286,0.6259193)
(0.9904762,0.6258667) (1.0000000,0.6258475)
};
\draw[angred,line width=.8pt,densely dashed] plot coordinates {
(0.0000000,0.2467667) (0.0190476,0.2268186) (0.0380952,0.2135848) (0.0571429,0.2000453)
(0.0761905,0.1887785) (0.0952381,0.1800180) (0.1142857,0.1709876) (0.1333333,0.1638715)
(0.1523810,0.1575049) (0.1714286,0.1512377) (0.1904762,0.1462852) (0.2095238,0.1414200)
(0.2285714,0.1369396) (0.2476190,0.1332015) (0.2666667,0.1293979) (0.2857143,0.1260856)
(0.3047619,0.1230948) (0.3238095,0.1200966) (0.3428571,0.1175425) (0.3619048,0.1150671)
(0.3809524,0.1126922) (0.4000000,0.1106304) (0.4190476,0.1085529) (0.4380952,0.1066520)
(0.4571429,0.1049196) (0.4761905,0.1031732) (0.4952381,0.1016202) (0.5142857,0.1001256)
(0.5333333,0.0986609) (0.5523810,0.0973556) (0.5714286,0.0960507) (0.5904762,0.0948208)
(0.6095238,0.0936912) (0.6285714,0.0925513) (0.6476190,0.0915084) (0.6666667,0.0905089)
(0.6857143,0.0895176) (0.7047619,0.0886169) (0.7238095,0.0877226) (0.7428571,0.0868628)
(0.7619048,0.0860678) (0.7809524,0.0852666) (0.8000000,0.0845182) (0.8190476,0.0838029)
(0.8380952,0.0830884) (0.8571429,0.0824293) (0.8761905,0.0817786) (0.8952381,0.0811443)
(0.9142857,0.0805540) (0.9333333,0.0799608) (0.9523810,0.0793977) (0.9714286,0.0788604)
(0.9904762,0.0783215) (1.0000000,0.0780653)
};
\end{scope}
\node[anchor=west] at (0,1.095) {$ -A/(a^2\omega_n^2) $};
\node[below] at (.5,-.14) {$n$};
\draw[angblue,line width=.9pt] (.48,.90)--(.56,.90);
\node[anchor=west] at (.58,.90) {$a=0.3$};
\draw[angred,line width=.9pt] (.48,.80)--(.56,.80);
\node[anchor=west] at (.58,.80) {$a=0.7$};
\node[anchor=west] at (.48,.45) {solid: $\mathrm{Re}$};
\node[anchor=west] at (.48,.34) {dashed: $\mathrm{Im}$};
\end{tikzpicture}%
}
\captionsetup{width=\linewidth,justification=justified,singlelinecheck=false}
\caption{Angular eigenvalues on the oblate QNM sequence. Solid and dashed curves show the real and imaginary parts of $-A/(a^2\omega_n^2)$, respectively; blue is $a=0.3$ and red is $a=0.7$. The limiting values are one and zero, confirming the quadratic oblate asymptotics evaluated at the QNM frequencies.}
\label{fig:thermal-angular-data}
\end{figure}

Next we turn to a computation of the QNM asymptotics to verify the above results directly.
At each value $a=0.3,0.7$, we have computed 251 oblate quasinormal modes, with $n=0,\ldots,250$.
Figure \ref{fig:thermal-angular-data} evaluates the ratio $-A/(a^2\omega_n^2)$ at the QNM frequencies. Its real part approaches one and its imaginary part approaches zero at both rotations.  

The frequencies approach $2\Omega_+-\ii\kappa_+(n-1)$. At the largest computed value of $n$, the imaginary frequency spacings at $a=0.3,0.7$ are $-0.2441389$ and $-0.2083037$ respectively, compared with $-\kappa_+=-0.2441067$ and $-0.2083090$ respectively showing good agreement.

We first consider fixed exterior radii $r>r'$, with $r'_*>\mathrm{Re}\,r_*(\ii a)$.
Using the up-up representation in \eqref{eq:three-residues}, the numerical data are compatible with the leading behaviour for the residues
\begin{equation}\label{eq:Eth-full}
\begin{aligned}
\underset{\omega=\omega_n^{\text{ob}}}{\mathrm{Res}}\widetilde G
&\sim\text{constant}\,n^{5/2}\exp\left\{\ii\omega_n^{\text{ob}}\left[r_*+r'_*-2r_*(\ii a)\right]\right\}.
\end{aligned}
\end{equation}
The corresponding leading sum, shifting the overtone number and using $\omega_n^{\text{ob}}\sim2\Omega_+-\ii\kappa_+n$, reads
\begin{widetext}
\begin{equation}\label{eq:thermal-sum}
\begin{aligned}
G_{\text{ob}}^{\mathrm{leading}}(t-t';r,r')&\propto e^{-2\ii\Omega_+(t-t')}\sum_{n=1}^{\infty}n^{5/2}\exp\left[-\kappa_+n\left(t-t'-r_*-r'_*+2r_*(\ii a)\right)\right]\\
&=e^{-2\ii\Omega_+(t-t')}\operatorname{Li}_{-5/2}\left(\exp\left[-\kappa_+\left(t-t'-r_*-r'_*+2r_*(\ii a)\right)\right]\right).
\end{aligned}
\end{equation}
\end{widetext}
The polylogarithm has branch point singularities when its argument reaches one, namely at
\begin{equation}\label{eq:th-bounces}
t-t'=r_*+r'_*-2r_*(\ii a)+\frac{2\pi k\ii}{\kappa_+},\qquad k\in\mathbb Z,
\end{equation}
which coincide with a subset of the candidate singularities from the geodesic analysis in \eqref{oblatesingsk}.
These form a discrete sequence on a vertical line in the complex $t-t'$ plane, with imaginary separation $2\pi/\kappa_+$. The corresponding open half-plane of absolute convergence is
\begin{equation}\label{eq:th-outer-convergence}
\mathrm{Re}\left(t-t'\right)>r_*+r'_*-2\mathrm{Re}\,r_*(\ii a).
\end{equation}

When both locations satisfy $r_+ < r,r' < r^\text{bounce}_\text{oblate}$ we consider the in-in representation of \eqref{eq:three-residues}.
The radial data support the approximation\footnote{Here we use a simplified asymptotic model in which we replace the oscillatory prefactor by an $n$-independent constant. This preserves the open half-plane of convergence.}
\begin{equation}\label{eq:inner-thermal-residue}
\begin{aligned}
\underset{\omega=\omega_n^{\text{ob}}}{\mathrm{Res}}\widetilde G
&=\frac{C_{\mathrm{up},\mathrm{in}}(\omega_n^{\text{ob}})R_{\mathrm{in},n}(r)R_{\mathrm{in},n}(r')}{\left.\frac{\dd W}{\dd\omega}\right|_{\omega_n^{\text{ob}}}}\\
&\sim\frac{\text{constant}}{n^{17/2}}\exp\left[\kappa_+n\left(2r_*(\ii a)-r_*-r'_*\right)\right].
\end{aligned}
\end{equation}
The corresponding mode sum reads
\begin{widetext}
\begin{equation}\label{eq:inner-thermal-sum}
\begin{aligned}
G_{\text{ob}}^{\mathrm{leading}}(t-t';r,r')&\propto e^{-2\ii\Omega_+(t-t')}\sum_{n=1}^{\infty}\frac{1}{n^{17/2}}\exp\left[-\kappa_+n\left(t-t'-2r_*(\ii a)+r_*+r'_*\right)\right]\\
&=e^{-2\ii\Omega_+(t-t')}\mathrm{Li}_{17/2}\left(\exp\left[-\kappa_+\left(t-t'-2r_*(\ii a)+r_*+r'_*\right)\right]\right).
\end{aligned}
\end{equation}
\end{widetext}
The polylogarithm branch points are at
\begin{equation}\label{eq:inner-thermal-bounces}
t-t'=2r_*(\ii a)-r_*-r'_*+\frac{2\pi k\ii}{\kappa_+},\qquad k\in\mathbb Z,
\end{equation}
which match a subset of the candidate singularities from the geodesic analysis in \eqref{oblatesingskpast}.
The corresponding open half-plane of absolute convergence is
\begin{equation}\label{eq:inner-thermal-region}
\mathrm{Re}\left(t-t'\right)>2\mathrm{Re}\,r_*(\ii a)-r_*-r'_*.
\end{equation}

For points on opposite sides of the oblate bounce, the up-in representation depends on the direct optical distance $r_*-r'_*$. The numerical results are compatible with the open convergence half-plane $\mathrm{Re}\left(t-t'\right)>r_*-r'_*$. Overall, combining the three configurations, we find the following region of convergence for the oblate family of QNMs:
\begin{equation}\label{eq:th-family-region}
\mathrm{Re}\left(t-t'\right)>|r_*-\mathrm{Re}\,r_*(\ii a)|+|r'_*-\mathrm{Re}\,r_*(\ii a)|.
\end{equation}

\section{Combined regions of convergence} \label{sec:combined}
At fixed $(s,\ell,m)$, the QNM contribution to the Green's function contains both the prolate and oblate families. Absolute convergence therefore requires both sums to converge, so the full region is the intersection of \eqref{eq:pr-family-region} and \eqref{eq:th-family-region}. For fixed exterior points, this gives
\begin{widetext}
\begin{equation}\label{eq:common-absolute-region}
\begin{aligned}
\mathrm{Re}\left(t-t'\right)>\max\Bigl\{|r_*-\mathrm{Re}\,r_*(\ii a)|+|r'_*-\mathrm{Re}\,r_*(\ii a)|,|x-\mathrm{Re}\,x(r_c)|+|x'-\mathrm{Re}\,x(r_c)|\Bigr\}.
\end{aligned}
\end{equation}
\end{widetext}
We remark again that these conditions are determined by null rays in the corresponding 2D optical metrics.

Figure \ref{fig:kerr-bounce-radii} shows how the two bounce radii depend on rotation. They approach the common Schwarzschild value $r^{\mathrm{bounce}}/M\simeq2.556929$ as $a\to0$, separate at nonzero rotation, and approach the common limit $1+\sqrt{2}$ as $a/M\to1^-$. Between these limits, the prolate bounce lies at larger radius than the oblate one.

For each fixed pair $r,r'$, the two individual convergence regions are right half-planes in the complex $t-t'$ plane. They are therefore nested, and their intersection is the half-plane to the right of the more restrictive boundary. The isolated singularities of the leading sums lie on the corresponding vertical lines. Changing the radial positions changes the real positions of these lines and can change which family fixes the common convergence boundary.

\begin{figure*}[t]
\centering
\begin{minipage}{0.65\textwidth}
\centering
\resizebox{\linewidth}{!}{%
\begin{tikzpicture}[x=24cm,y=.055cm,every node/.style={font=\footnotesize},>=Stealth]
\fill[green!10] (5.481912,-50) rectangle (5.70,50);
\draw[->,gray!80] (5.30,-51)--(5.30,55) node[above] {$\mathrm{Im}\left(t-t'\right)$};
\draw[->,gray!80] (5.30,0)--(5.71,0) node[right] {$\mathrm{Re}\left(t-t'\right)$};
\foreach \yy in {-50,-25,0,25,50} {\draw[gray!70] (5.297,\yy)--(5.303,\yy);\node[anchor=east] at (5.294,\yy) {\scriptsize $\yy$};}
\draw[blue!75!black,thick,dashed] (5.352306,-50)--(5.352306,50);
\draw[red!75!black,thick,dashed] (5.481912,-50)--(5.481912,50);
\node[blue!75!black] at (5.352306,58) {prolate};
\node[red!75!black] at (5.481912,58) {oblate};
\foreach \yy in {-36.907297,-12.302432,12.302432,36.907297} {\fill[blue!75!black] (5.352306,\yy) circle (2.1pt);}
\foreach \yy in {-43.215149,-13.052340,17.110469,47.273278} {\draw[red!75!black,line width=.7pt,fill=white] (5.481912,\yy) circle (2.4pt);}
\node[green!35!black,align=center,text width=3.3cm] at (5.595,29) {absolute convergence\\of both families};
\node[blue!75!black,anchor=north,align=center] at (5.352306,-55) {$5.352306$};
\node[red!75!black,anchor=north,align=center] at (5.481912,-55) {$5.481912$};
\node at (5.50,72) {$a=0.7,\qquad (r,r')=(4,3)$};
\end{tikzpicture}
}
\captionsetup{
  width=\linewidth,
  justification=justified,
  singlelinecheck=false
}
\caption{An enlargement of the complex $t-t'$ plane near the two outer convergence boundaries. Filled blue points are logarithmic prolate branch points; open red points are oblate branch points of the $\mathrm{Li}_{-5/2}$ model. The common open half-plane of absolute convergence is shaded green. The dashed lines locate the convergence boundaries.}
\label{fig:outer-lines}
\end{minipage}
\end{figure*}
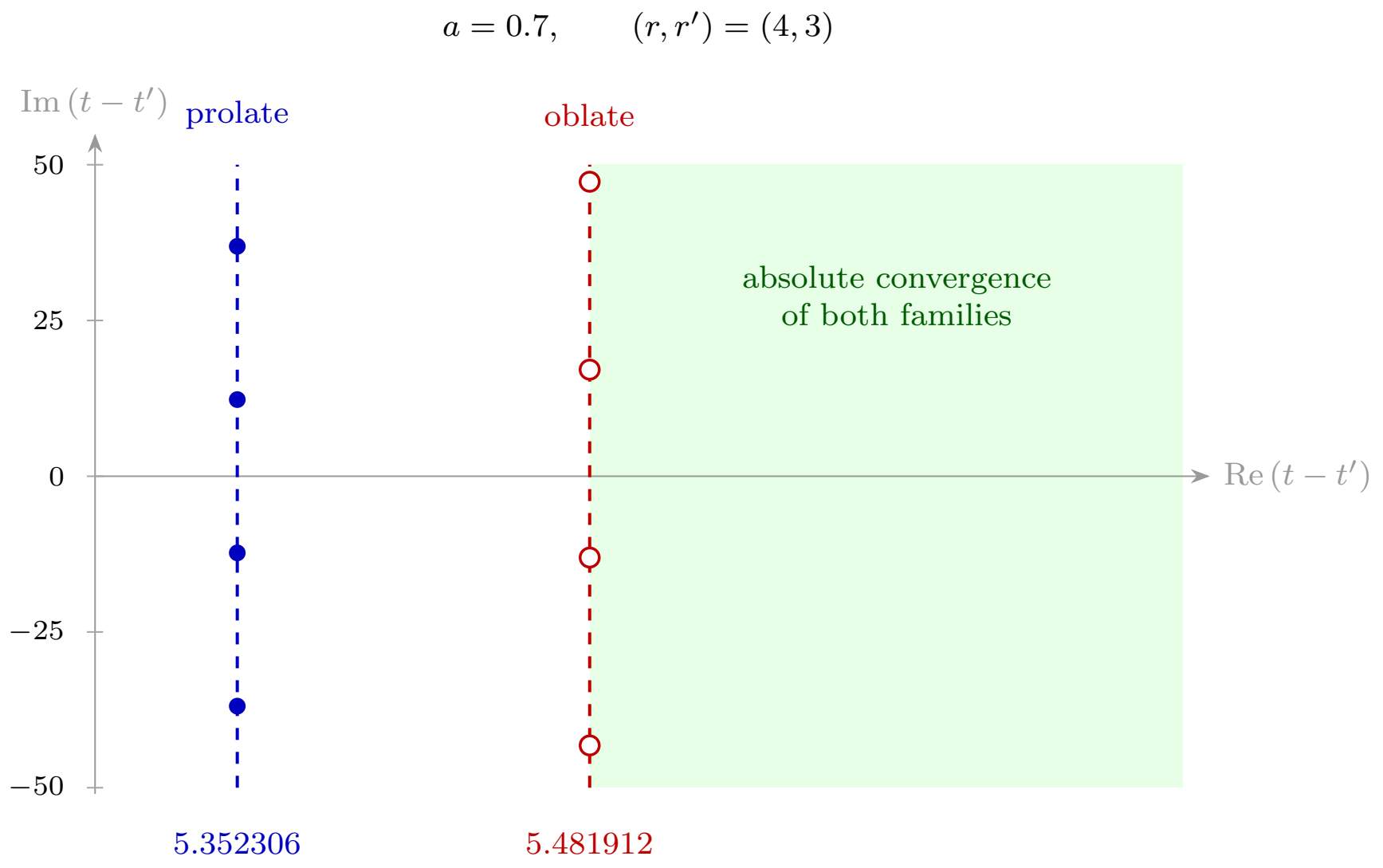

Figure \ref{fig:outer-lines} illustrates a configuration where $r,r' > r^\text{bounce}_\text{prolate}, r^\text{bounce}_\text{oblate}$, at $a=0.7$, with $(r,r')=(4,3)$. The prolate and oblate boundaries are at $\mathrm{Re}\left(t-t'\right)=5.352306$ and $\mathrm{Re}\left(t-t'\right)=5.481912$, respectively. The oblate line lies farther to the right, so the common open region of absolute convergence is $\mathrm{Re}\left(t-t'\right)>5.481912$.

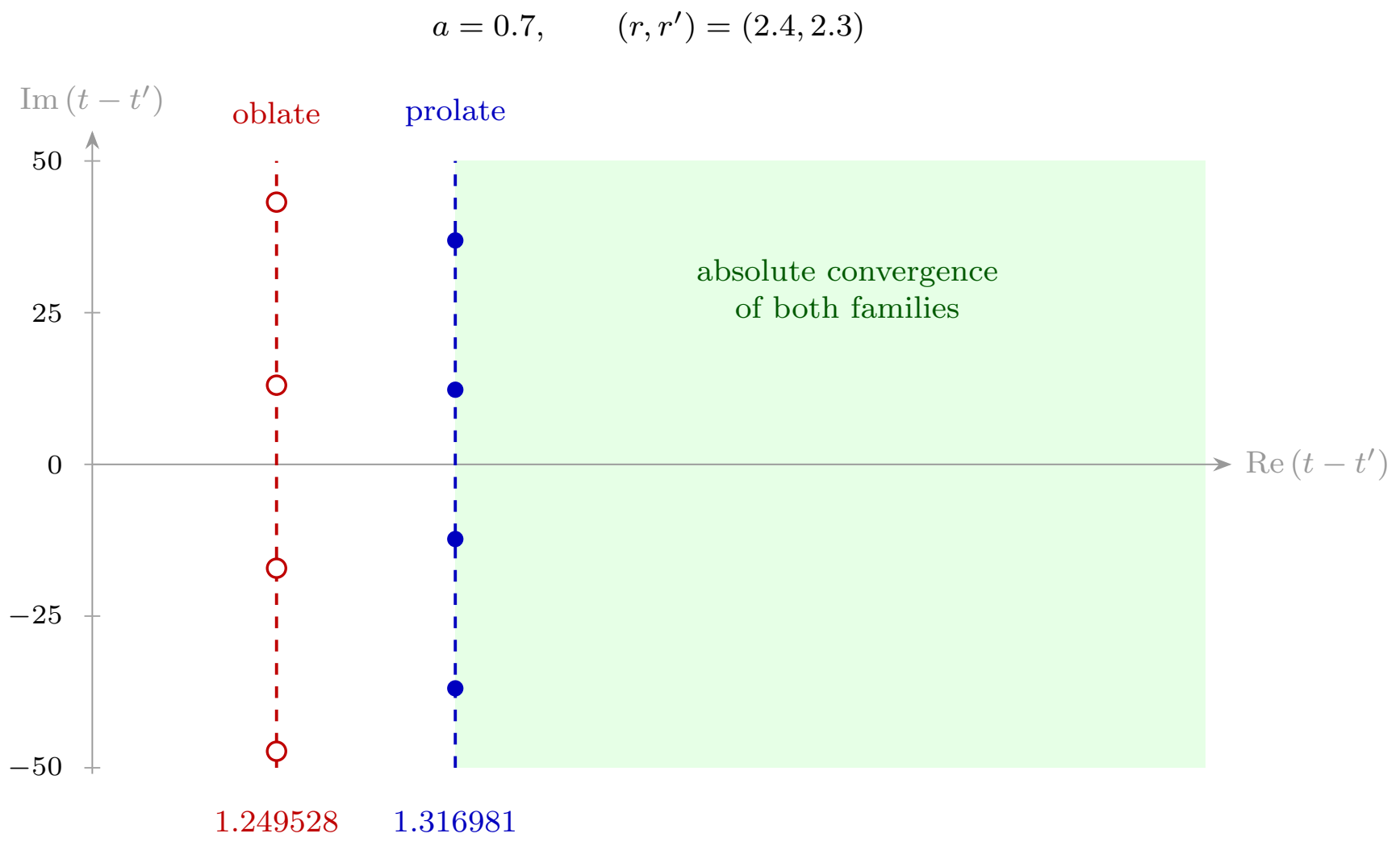
\begin{figure*}[t]
\centering
\begin{minipage}{0.65\textwidth}
\centering
\resizebox{\linewidth}{!}{%
\begin{tikzpicture}[x=24cm,y=.055cm,every node/.style={font=\footnotesize},>=Stealth]
\fill[green!10] (1.316981,-50) rectangle (1.60,50);
\draw[->,gray!80] (1.18,-51)--(1.18,55) node[above] {$\mathrm{Im}\left(t-t'\right)$};
\draw[->,gray!80] (1.18,0)--(1.61,0) node[right] {$\mathrm{Re}\left(t-t'\right)$};
\foreach \yy in {-50,-25,0,25,50} {
  \draw[gray!70] (1.177,\yy)--(1.183,\yy);
  \node[anchor=east] at (1.174,\yy) {\scriptsize $\yy$};
}
\draw[red!75!black,thick,dashed] (1.249528,-50)--(1.249528,50);
\draw[blue!75!black,thick,dashed] (1.316981,-50)--(1.316981,50);
\node[red!75!black] at (1.249528,58) {oblate};
\node[blue!75!black] at (1.316981,58) {prolate};
\foreach \yy in {-36.907297,-12.302432,12.302432,36.907297} {
  \fill[blue!75!black] (1.316981,\yy) circle (2.1pt);
}
\foreach \yy in {-47.273278,-17.110469,13.052340,43.215149} {
  \draw[red!75!black,line width=.7pt,fill=white] (1.249528,\yy) circle (2.4pt);
}
\node[green!35!black,align=center,text width=4cm] at (1.465,29) {absolute convergence\\of both families};
\node[red!75!black,anchor=north] at (1.249528,-55) {$1.249528$};
\node[blue!75!black,anchor=north] at (1.316981,-55) {$1.316981$};
\node at (1.39,72) {$a=0.7,\qquad (r,r')=(2.4,2.3)$};
\end{tikzpicture}
}
\captionsetup{
  width=\linewidth,
  justification=justified,
  singlelinecheck=false
}
\caption{An inner configuration, with both $r,r'$ between the bouncing radii and the outer horizon. Filled blue points are logarithmic prolate branch points. Open red points are oblate branch points of the $\mathrm{Li}_{17/2}$ model. The prolate line is farther to the right and fixes the common open half-plane of absolute convergence, shaded green.}
\label{fig:inner-lines}
\end{minipage}
\end{figure*}

Figure \ref{fig:inner-lines} shows $r,r' < r^\text{bounce}_\text{prolate}, r^\text{bounce}_\text{oblate}$ for $(r,r')=(2.4,2.3)$ at the same rotation $a=0.7$. Both points are outside the horizon. Here, the oblate boundary is at $\mathrm{Re}\left(t-t'\right)=1.249528$, while the prolate boundary is at $\mathrm{Re}\left(t-t'\right)=1.316981$. The prolate family therefore fixes the common open region, $\mathrm{Re}\left(t-t'\right)>1.316981$.

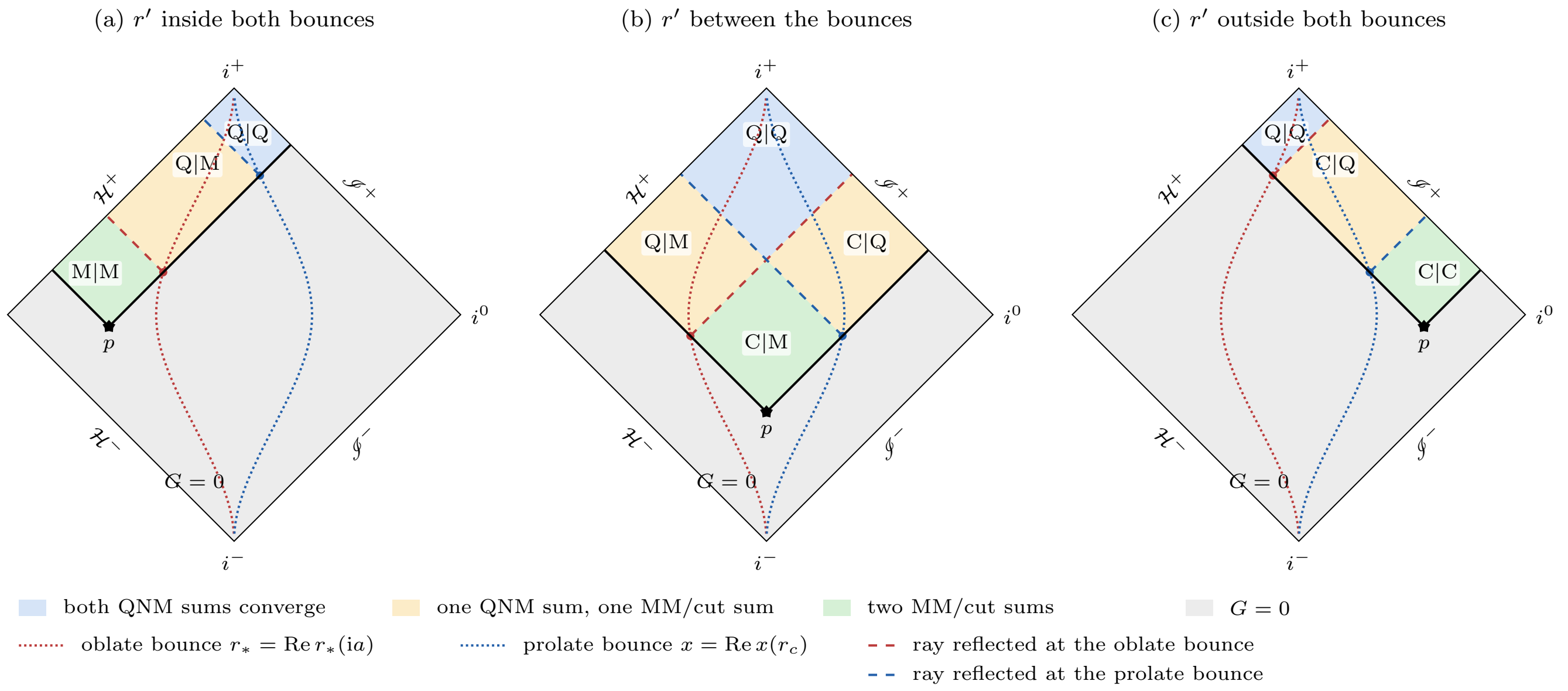
\begin{figure*}[t]
\centering
\begin{minipage}{\textwidth}
\centering
\resizebox{\linewidth}{!}{%
\providecommand{\NC}[2]{({(atan(#2)-atan(#1))/180},{(atan(#2)+atan(#1))/180})}
\def\BIG{9000}
\begin{tikzpicture}[x=2.75cm,y=2.75cm,every node/.style={font=\scriptsize},>=Stealth,reg/.style={fill=white,fill opacity=.75,text opacity=1,inner sep=.8pt,rounded corners=1pt}]
\definecolor{cQQ}{RGB}{214,228,247}  
\definecolor{cMX}{RGB}{252,236,200}  
\definecolor{cMM}{RGB}{214,240,214}  
\definecolor{cZZ}{RGB}{236,236,236}  
\definecolor{bth}{RGB}{187,60,60}    
\definecolor{bpr}{RGB}{36,96,170}    
\def\bth{-0.6}\def\bpr{0.6}          
\foreach \panel/\sh/\tp/\sp/\ttl in {0/0/-0.2/-1.2/{(a) $r'$ inside both bounces},
                                      1/2.35/-0.8/0/{(b) $r'$ between the bounces},
                                      2/4.70/-0.2/1.2/{(c) $r'$ outside both bounces}} {
\begin{scope}[shift={(\sh,0)}]
  \pgfmathsetmacro{\up}{\tp-\sp}\pgfmathsetmacro{\vp}{\tp+\sp}
  \fill[cZZ] (-1,0)--(0,1)--(1,0)--(0,-1)--cycle;
  \ifnum\panel=0
    \pgfmathsetmacro{\vA}{\up+2*\bth}\pgfmathsetmacro{\vC}{\up+2*\bpr}
    \fill[cMM] \NC{\up}{\vp}--\NC{\up}{\vA}--\NC{\BIG}{\vA}--\NC{\BIG}{\vp}--cycle;
    \fill[cMX] \NC{\up}{\vA}--\NC{\up}{\vC}--\NC{\BIG}{\vC}--\NC{\BIG}{\vA}--cycle;
    \fill[cQQ] \NC{\up}{\vC}--\NC{\up}{\BIG}--(0,1)--\NC{\BIG}{\vC}--cycle;
    \draw[bth,thick,dashed] \NC{\up}{\vA}--\NC{\BIG}{\vA};
    \draw[bpr,thick,dashed] \NC{\up}{\vC}--\NC{\BIG}{\vC};
    \fill[bth] \NC{\up}{\vA} circle (.018); \fill[bpr] \NC{\up}{\vC} circle (.018);
    \node[reg] at \NC{3}{-0.8} {M$|$M};
    \node[reg] at \NC{3.5}{1.0} {Q$|$M};
    \node[reg] at \NC{2.3}{4.5} {Q$|$Q};
  \fi
  \ifnum\panel=1
    \pgfmathsetmacro{\uA}{\vp-2*\bth}\pgfmathsetmacro{\vC}{\up+2*\bpr}
    \fill[cMM] \NC{\up}{\vp}--\NC{\uA}{\vp}--\NC{\uA}{\vC}--\NC{\up}{\vC}--cycle;
    \fill[cMX] \NC{\uA}{\vp}--\NC{\BIG}{\vp}--\NC{\BIG}{\vC}--\NC{\uA}{\vC}--cycle;
    \fill[cMX] \NC{\up}{\vC}--\NC{\uA}{\vC}--\NC{\uA}{\BIG}--\NC{\up}{\BIG}--cycle;
    \fill[cQQ] \NC{\uA}{\vC}--\NC{\BIG}{\vC}--(0,1)--\NC{\uA}{\BIG}--cycle;
    \draw[bth,thick,dashed] \NC{\uA}{\vp}--\NC{\uA}{\BIG};
    \draw[bpr,thick,dashed] \NC{\up}{\vC}--\NC{\BIG}{\vC};
    \fill[bth] \NC{\uA}{\vp} circle (.018); \fill[bpr] \NC{\up}{\vC} circle (.018);
    \node[reg] at \NC{-0.2}{-0.2} {C$|$M};
    \node[reg] at \NC{2.5}{-0.2} {Q$|$M};
    \node[reg] at \NC{-0.2}{2.5} {C$|$Q};
    \node[reg] at \NC{3}{3} {Q$|$Q};
  \fi
  \ifnum\panel=2
    \pgfmathsetmacro{\uC}{\vp-2*\bpr}\pgfmathsetmacro{\uA}{\vp-2*\bth}
    \fill[cMM] \NC{\up}{\vp}--\NC{\uC}{\vp}--\NC{\uC}{\BIG}--\NC{\up}{\BIG}--cycle;
    \fill[cMX] \NC{\uC}{\vp}--\NC{\uA}{\vp}--\NC{\uA}{\BIG}--\NC{\uC}{\BIG}--cycle;
    \fill[cQQ] \NC{\uA}{\vp}--\NC{\BIG}{\vp}--(0,1)--\NC{\uA}{\BIG}--cycle;
    \draw[bpr,thick,dashed] \NC{\uC}{\vp}--\NC{\uC}{\BIG};
    \draw[bth,thick,dashed] \NC{\uA}{\vp}--\NC{\uA}{\BIG};
    \fill[bpr] \NC{\uC}{\vp} circle (.018); \fill[bth] \NC{\uA}{\vp} circle (.018);
    \node[reg] at \NC{-0.8}{3} {C$|$C};
    \node[reg] at \NC{1.0}{3.5} {C$|$Q};
    \node[reg] at \NC{4.5}{2.3} {Q$|$Q};
  \fi
  \draw[bth,densely dotted,thick] plot[domain=-87:87,samples=120,variable=\th]
    ({(atan(tan(\th)+\bth)-atan(tan(\th)-\bth))/180},{(atan(tan(\th)+\bth)+atan(tan(\th)-\bth))/180});
  \draw[bpr,densely dotted,thick] plot[domain=-87:87,samples=120,variable=\th]
    ({(atan(tan(\th)+\bpr)-atan(tan(\th)-\bpr))/180},{(atan(tan(\th)+\bpr)+atan(tan(\th)-\bpr))/180});
  \draw[black,thick] \NC{\up}{\vp}--\NC{\up}{\BIG};
  \draw[black,thick] \NC{\up}{\vp}--\NC{\BIG}{\vp};
  \node[star,star points=5,fill=black,inner sep=1.1pt] at \NC{\up}{\vp} {};
  \node[below=1pt] at \NC{\up}{\vp} {$p$};
  \node at \NC{-1.2}{-7} {$G=0$};
  \draw (-1,0)--(0,1)--(1,0)--(0,-1)--cycle;
  \node[rotate=45,above] at (-.5,.5) {$\mathcal H^+$};
  \node[rotate=-45,below] at (-.5,-.5) {$\mathcal H^-$};
  \node[rotate=-45,above] at (.5,.5) {$\mathscr I^+$};
  \node[rotate=45,below] at (.5,-.5) {$\mathscr I^-$};
  \node[above] at (0,1) {$i^+$};\node[right] at (1,0) {$i^0$};\node[below] at (0,-1) {$i^-$};
  \node[font=\footnotesize] at (0,1.3) {\ttl};
\end{scope}
}
\begin{scope}[shift={(-0.95,-1.33)}]
  \fill[cQQ] (0,0) rectangle (.12,.07); \node[anchor=west] at (.15,.035) {both QNM sums converge};
  \fill[cMX] (1.65,0) rectangle (1.77,.07); \node[anchor=west] at (1.8,.035) {one QNM sum, one MM/cut sum};
  \fill[cMM] (3.55,0) rectangle (3.67,.07); \node[anchor=west] at (3.7,.035) {two MM/cut sums};
  \fill[cZZ] (5.15,0) rectangle (5.27,.07); \node[anchor=west] at (5.3,.035) {$G=0$};
  \draw[bth,densely dotted,thick] (0,-.13)--(.2,-.13); \node[anchor=west] at (.23,-.13) {oblate bounce $r_*=\mathrm{Re}\,r_*(\ii a)$};
  \draw[bpr,densely dotted,thick] (1.95,-.13)--(2.15,-.13); \node[anchor=west] at (2.18,-.13) {prolate bounce $x=\mathrm{Re}\,x(r_c)$};
  \draw[bth,thick,dashed] (3.75,-.13)--(3.87,-.13); \node[anchor=west] at (3.9,-.13) {ray reflected at the oblate bounce};
  \draw[bpr,thick,dashed] (3.75,-.26)--(3.87,-.26); \node[anchor=west] at (3.9,-.26) {ray reflected at the prolate bounce};
\end{scope}
\end{tikzpicture}
}
\captionsetup{
  width=\linewidth,
  justification=justified,
  singlelinecheck=false
}
\caption{Convergence regions of the QNM sums at fixed $(\ell,m)$ in the Kerr exterior, for a source at $p=(t',r')$. The dotted curves are the two bounce radii, prolate (blue) and oblate (red); the prolate one lies at larger radius for all $0<a<M$ (FIG.~\ref{fig:kerr-bounce-radii}), and their separation is exaggerated. Solid black lines are the forward light cone of $p$, outside of which $G=0$. Dashed lines are rays reflected at a bounce radius. Each region is labelled by the representation of the oblate and prolate contributions in an ordered pair. Q denotes a convergent QNM sum; M denotes the expected convergent sum of horizon Matsubara modes, $\omega-m\Omega_+\in\ii\kappa_+\mathbb{Z}$, for regions reaching $\mathcal H^+$; C denotes the cut contribution at null infinity, which can be thought of as arising from a set of Matsubara poles in the Kerr-de Sitter solution coalescing into the branch cut at $\omega=0$ as $\Lambda\to 0^+$. \textbf{(a)} $r'$ below both bounces: both reflected rays are ingoing, and the oblate one comes first. \textbf{(b)} $r'$ between the bounces: the ingoing ray reflects at the oblate bounce and the outgoing ray at the prolate one, leaving four regions. \textbf{(c)} $r'$ above both bounces: both reflected rays are outgoing, and the prolate one comes first. For $a=0.7$ these orderings agree with the convergence boundaries in FIGs.~\ref{fig:outer-lines} and \ref{fig:inner-lines}. For simplicity the diagram is an approximation where all rays are drawn at $45^\circ$. More accurately, the oblate boundaries are null with respect to $r_*$ and the prolate ones with respect to $x$.}
\label{fig:kerr-regions}
\end{minipage}
\end{figure*}

The identified convergence region for QNMs determines a region decomposition for the Green's function in terms of Matsubara modes and cut contributions as discussed for Schwarzschild(-de Sitter) in \cite{Arnaudo:2025uos}.
In Figure \ref{fig:kerr-regions}, we organise the convergence regions in a schematic spacetime picture for a source at $p=(t',r')$. 
Recall the prolate and oblate optical metrics have different null lines, so it is not possible to draw a single conformal diagram where both boundaries of convergence are straight lines at 45 degrees.  The two optical coordinates $x, r_*$ are nevertheless close in the real exterior. Indeed,
\begin{equation}
\frac{\dd x}{\dd r_*}=\sqrt{1-\frac{a^2\Delta}{(r^2+a^2)^2}},
\qquad
\frac{\sqrt{1+2\sqrt{2}}}{2}\leq\frac{\dd x}{\dd r_*}\le 1.
\end{equation}
Thus the prolate characteristics, $\dd t/\dd r_*=\pm\dd x/\dd r_*$, differ in slope from the oblate characteristics, $\dd t/\dd r_*=\pm1$, by at most approximately $2.17\%$ throughout the exterior. With our normalization $x-r_*\to0$ at infinity, their difference is positive, remains finite at the outer horizon, and decreases towards zero at infinity. For visual clarity, we will draw the prolate characteristics as if they were also null rays of the oblate diagram. This is an approximation for simplifying the presentation; the strict criterion is given in \eqref{eq:common-absolute-region}.

The three panels in Figure \ref{fig:kerr-regions} explore three configurations for the source placement: to the left of both bounces, between them, and to the right of both. Close to the outer horizon, we expect a completion in the mode decomposition provided by Matsubara modes. More precisely, when a family of QNMs fails to converge, we expect the corresponding contribution to be replaced by a Matsubara mode sum. This would lead (for example referring to the left panel) to a region described entirely in terms of Matsubara modes, and an intermediate region where a reconstruction is provided in terms of a mix of oblate QNMs, and a Matsubara contribution replacing the diverging prolate QNM sum. Close to infinity, the same situation appears, but the sum in terms of Matsubara modes is replaced by cut contributions due to the presence of the irregular singularity. The cut contributions can be thought of as arising from a set of Matsubara poles in the Kerr-de Sitter solution coalescing into the branch cut at $\omega=0$ as $\Lambda\to 0^+$.

\section{Discussion} \label{sec:discussion}

The Schwarzschild limit provides a useful check of the two optical descriptions. Restoring $M$, both optical metrics reduce as $a\to0$ to
\begin{equation}
\dd s^2_{\mathrm{pr}},\dd s^2_{\mathrm{oblate}}\to-\left(1-\frac{2M}{r}\right)\dd t^2+\frac{\dd r^2}{1-2M/r}.
\end{equation}
With the normalization adopted above, their optical coordinates therefore coincide in this limit:
\begin{equation}
x(r)\to r_*(r)=r+2M\log\left(\frac{r}{2M}-1\right).
\end{equation}
The complex points $r_c,\bar r_c$ and $\ii a$ approach $r=0$. 
The optical separation entering \eqref{eq:quoted-prolate-spacing} consequently approaches $4\pi\ii M$, and both positive spacings recover the Schwarzschild value:
\begin{equation}
\rho,\kappa_+\to\frac{1}{4M}=\frac{2\pi}{\beta},
\end{equation}
consistently with the discussion in \cite{Arnaudo:2026tcy}.

The real parts of the complex optical values vanish, so both definitions of the exterior bounce radius reduce to $r_*(r^{\mathrm{bounce}})=0$, giving 
\begin{equation}
r^{\mathrm{bounce}}=2M\left[1+W_0(e^{-1})\right]\simeq2.556929M.
\end{equation}
The two outer sequences of complex-time singularities also approach the same lattice,
\begin{equation}
t-t'=r_*+r'_*+\ii\beta\left(k+\frac12\right),\qquad k\in\mathbb Z.
\end{equation}

We underline that the regional structure analysed in the previous section becomes important when considering perturbations generated by sources. The sourced response is obtained by convolving the retarded Green's function with the source, and each source event $(t',r')$ carries its own convergence regions. At a fixed observation point $(t,r)$, different portions of the source history can therefore contribute through different representations of the Green's function. The source integral should accordingly be partitioned into domains where the appropriate QNM, Matsubara or cut representations apply. This is analogous to the mixed mode decompositions arising from extended initial data discussed in \cite{Arnaudo:2025uos}.

For example, for a plunging particle, the source follows a trajectory $r'=r'(t')$, and the convergence conditions must be evaluated along this trajectory. The subdivision depends jointly on the emission radius, the observation radius and the elapsed time $t-t'$. In Kerr, the two bounce radii introduce a further distinction: a given segment of the source history may admit a convergent QNM representation for both families, for only one family, or for neither. Consequently, the response at a single observation time can combine QNM contributions from some portions of the trajectory with Matsubara or cut contributions from others. This was recently studied in \cite{Oshita:2025qmn, Su:2026gmp}.

Kerr has an inner horizon and two families of QNMs with different optical metrics. It is instructive to compare this case to other black holes with an inner horizon. For this purpose, for the remainder of this discussion we turn to a massless scalar of charge $q$ on a Reissner--Nordstr\"om black hole of charge $Q$. We find that it has only one optical metric. Its radial equation is
\begin{equation}\label{eq:RN-radial}
\frac{\dd}{\dd r}\left(\Delta\frac{\dd R}{\dd r}\right)+\left[\frac{(\omega r^2-qQr)^2}{\Delta}-\ell(\ell+1)\right]R=0,
\end{equation}
with $\Delta=r^2-2Mr+Q^2=(r-r_+)(r-r_-)$ and $\kappa_{\pm}=(r_+-r_-)/(2r_\pm^2)$.
At large frequency the characteristics are $\dd t/\dd r=\pm r^2/\Delta$ for both signs of $\mathrm{Re}\,\omega$. There is therefore one optical metric, $\dd s^2=-f\,\dd t^2+f^{-1}\dd r^2$ with $f=\Delta/r^2$.
The presence of the scalar charge $q$ does not change the optical geometry. 

\begin{figure}[h!]
\centering
\includegraphics[width=\linewidth]{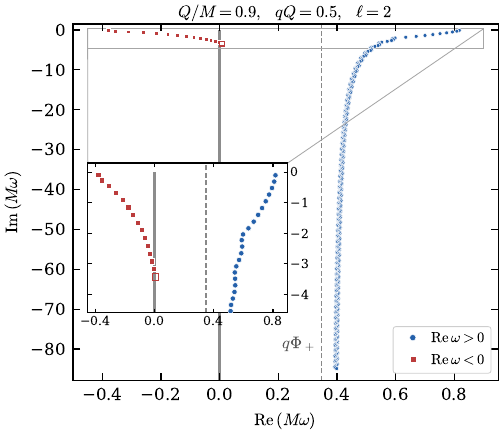}
\captionsetup{width=\linewidth,justification=justified,singlelinecheck=false}
\caption{Quasinormal frequencies of a massless charged scalar on Reissner--Nordstr\"om, for $Q/M=0.9$, $qQ=0.5$, $\ell=2$. Blue circles: the $402$ modes with $\mathrm{Re}\,\omega>0$, which form a single vertical line with spacing $\kappa_+$. Red squares: the modes with $\mathrm{Re}\,\omega<0$. They reach the branch cut on the negative imaginary axis (thick gray line) after fourteen modes; the open square is the first pole behind the cut. The dashed line marks $\mathrm{Re}\,\omega=qQ/r_+$. The inset enlarges $|\mathrm{Im}\,M\omega|<4.6$.}
\label{fig:RN-qnm-09}
\end{figure}

\begin{figure}[h!]
\centering
\includegraphics[width=\linewidth]{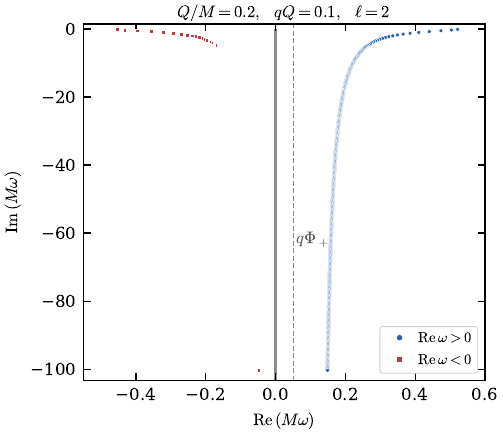}
\captionsetup{width=\linewidth,justification=justified,singlelinecheck=false}
\caption{As in Figure \ref{fig:RN-qnm-09}, for $Q/M=0.2$, $qQ=0.1$. Both families survive, with $402$ modes each, and form two vertical lines with spacing $\kappa_+$. The lines are approximately symmetric about $\mathrm{Re}\,\omega=qQ/r_+$ (dashed).}
\label{fig:RN-qnm-02}
\end{figure}

Using the same numerical methods as in Kerr, we tracked two pole sequences at $\ell=2$ for $(Q/M,qQ)=(0.9,0.5)$ and $(0.2,0.1)$, computing up to approximately $400$ poles per sequence. At low overtones, the two sequences have different sign of the real part; see Figures \ref{fig:RN-qnm-09} and \ref{fig:RN-qnm-02}.

At $Q/M=0.2$, both sequences persist on the physical sheet and approach vertical lines, each with mean spacing $\kappa_+$, where $\kappa_+M=0.24997$. The two lines are approximately symmetric about $\mathrm{Re}\,\omega=qQ/r_+$.

At $Q/M=0.9$, only one sequence develops into a vertical line on the physical sheet. Its mean spacing is $0.21141/M$, in agreement with $\kappa_+=0.211414/M$, with individual spacings modulated on a period $\kappa_-$ along the imaginary-frequency axis. The other sequence contains fourteen poles with negative real frequency before reaching the negative imaginary axis at $|\mathrm{Im}(M\omega)|\simeq3.4$ and continuing behind the branch cut. 

The residues follow the same pattern as the oblate Kerr family, with $r=\ii a$ replaced by $r=0$:
\begin{equation}\label{eq:RN-residue}
\underset{\omega=\omega_n}{\mathrm{Res}}\,\widetilde G\sim\text{constant}\,\frac{1}{n}\,\exp\left\{\ii\omega_n\left[r_*+r'_*-2r_*(0)\right]\right\}.
\end{equation}
By defining
\be
r_*(r)=r+\frac{r_+^2}{r_+-r_-}\log\frac{r-r_+}{r_+}-\frac{r_-^2}{r_+-r_-}\log\frac{r-r_-}{r_+},
\ee
we have $r_*(0)=U+2\pi\ii M$, where
\begin{equation}
U=\frac{r_-^2}{r_+-r_-}\log\frac{r_+}{r_-},
\end{equation}
and $r_*$ is continued from the exterior through the upper half plane. The bounce radius is therefore defined by $r_*(r^{\mathrm{bounce}})=U$. 
In Reissner--Nordstr\"om the bounce at $r=0$ has a real geometric counterpart. Radial timelike geodesics have a turning point that in the optical limit approaches $r=0$, so that the bouncing null curve is the limit of real timelike geodesics reflected by the timelike singularity.

\renewcommand{\addcontentsline}[3]{}
\paragraph{Acknowledgements.}
\begin{acknowledgments}
We thank Luis Lehner for discussions.
PA is supported by the Royal Society grant URF\textbackslash R\textbackslash 231002, `Dynamics of holographic field theories'.
BW is supported by a Royal Society University Research Fellowship and in part by the Science and Technology Facilities Council (Consolidated Grant `New Frontiers in Particle Physics, Cosmology and Gravity'). 
\end{acknowledgments}

		\bibliography{refs}

\begin{thebibliography}{30}%
\makeatletter
\providecommand \@ifxundefined [1]{%
 \@ifx{#1\undefined}
}%
\providecommand \@ifnum [1]{%
 \ifnum #1\expandafter \@firstoftwo
 \else \expandafter \@secondoftwo
 \fi
}%
\providecommand \@ifx [1]{%
 \ifx #1\expandafter \@firstoftwo
 \else \expandafter \@secondoftwo
 \fi
}%
\providecommand \natexlab [1]{#1}%
\providecommand \enquote  [1]{``#1''}%
\providecommand \bibnamefont  [1]{#1}%
\providecommand \bibfnamefont [1]{#1}%
\providecommand \citenamefont [1]{#1}%
\providecommand \href@noop [0]{\@secondoftwo}%
\providecommand \href [0]{\begingroup \@sanitize@url \@href}%
\providecommand \@href[1]{\@@startlink{#1}\@@href}%
\providecommand \@@href[1]{\endgroup#1\@@endlink}%
\providecommand \@sanitize@url [0]{\catcode `\\12\catcode `\$12\catcode `\&12\catcode `\#12\catcode `\^12\catcode `\_12\catcode `\%12\relax}%
\providecommand \@@startlink[1]{}%
\providecommand \@@endlink[0]{}%
\providecommand \url  [0]{\begingroup\@sanitize@url \@url }%
\providecommand \@url [1]{\endgroup\@href {#1}{\urlprefix }}%
\providecommand \urlprefix  [0]{URL }%
\providecommand \Eprint [0]{\href }%
\providecommand \doibase [0]{https://doi.org/}%
\providecommand \selectlanguage [0]{\@gobble}%
\providecommand \bibinfo  [0]{\@secondoftwo}%
\providecommand \bibfield  [0]{\@secondoftwo}%
\providecommand \translation [1]{[#1]}%
\providecommand \BibitemOpen [0]{}%
\providecommand \bibitemStop [0]{}%
\providecommand \bibitemNoStop [0]{.\EOS\space}%
\providecommand \EOS [0]{\spacefactor3000\relax}%
\providecommand \BibitemShut  [1]{\csname bibitem#1\endcsname}%
\let\auto@bib@innerbib\@empty
\bibitem [{\citenamefont {Abac}\ \emph {et~al.}(2026)\citenamefont {Abac} \emph {et~al.}}]{LVKloud}%
  \BibitemOpen
  \bibfield  {author} {\bibinfo {author} {\bibfnamefont {A.~G.}\ \bibnamefont {Abac}} \emph {et~al.} (\bibinfo {collaboration} {The LIGO Scientific Collaboration, The Virgo Collaboration, and The KAGRA Collaboration}),\ }\bibfield  {title} {\bibinfo {title} {{Black Hole Spectroscopy and Tests of General Relativity with GW250114}},\ }\href {https://doi.org/10.1103/6c61-fm1n} {\bibfield  {journal} {\bibinfo  {journal} {Phys. Rev. Lett.}\ }\textbf {\bibinfo {volume} {136}},\ \bibinfo {pages} {041403} (\bibinfo {year} {2026})},\ \Eprint {https://arxiv.org/abs/2509.08099} {arXiv:2509.08099 [gr-qc]} \BibitemShut {NoStop}%
\bibitem [{\citenamefont {Berti}\ \emph {et~al.}(2025)\citenamefont {Berti} \emph {et~al.}}]{Berti:2025hly}%
  \BibitemOpen
  \bibfield  {author} {\bibinfo {author} {\bibfnamefont {E.}~\bibnamefont {Berti}} \emph {et~al.},\ }\bibfield  {title} {\bibinfo {title} {{Black hole spectroscopy: from theory to experiment}},\ }\href@noop {} {\  (\bibinfo {year} {2025})},\ \Eprint {https://arxiv.org/abs/2505.23895} {arXiv:2505.23895 [gr-qc]} \BibitemShut {NoStop}%
\bibitem [{\citenamefont {Leaver}(1986)}]{Leaver}%
  \BibitemOpen
  \bibfield  {author} {\bibinfo {author} {\bibfnamefont {E.~W.}\ \bibnamefont {Leaver}},\ }\bibfield  {title} {\bibinfo {title} {Spectral decomposition of the perturbation response of the schwarzschild geometry},\ }\href {https://doi.org/10.1103/PhysRevD.34.384} {\bibfield  {journal} {\bibinfo  {journal} {Phys. Rev. D}\ }\textbf {\bibinfo {volume} {34}},\ \bibinfo {pages} {384} (\bibinfo {year} {1986})}\BibitemShut {NoStop}%
\bibitem [{\citenamefont {Andersson}(1997)}]{Andersson:1996cm}%
  \BibitemOpen
  \bibfield  {author} {\bibinfo {author} {\bibfnamefont {N.}~\bibnamefont {Andersson}},\ }\bibfield  {title} {\bibinfo {title} {{Evolving test fields in a black hole geometry}},\ }\href {https://doi.org/10.1103/PhysRevD.55.468} {\bibfield  {journal} {\bibinfo  {journal} {Phys. Rev. D}\ }\textbf {\bibinfo {volume} {55}},\ \bibinfo {pages} {468} (\bibinfo {year} {1997})},\ \Eprint {https://arxiv.org/abs/gr-qc/9607064} {arXiv:gr-qc/9607064} \BibitemShut {NoStop}%
\bibitem [{\citenamefont {Kuntz}(2026)}]{Kuntz:2025gdq}%
  \BibitemOpen
  \bibfield  {author} {\bibinfo {author} {\bibfnamefont {A.}~\bibnamefont {Kuntz}},\ }\bibfield  {title} {\bibinfo {title} {{Green function of the P{\"o}schl-Teller potential}},\ }\href {https://doi.org/10.21468/SciPostPhys.20.4.120} {\bibfield  {journal} {\bibinfo  {journal} {SciPost Phys.}\ }\textbf {\bibinfo {volume} {20}},\ \bibinfo {pages} {120} (\bibinfo {year} {2026})},\ \Eprint {https://arxiv.org/abs/2510.17954} {arXiv:2510.17954 [gr-qc]} \BibitemShut {NoStop}%
\bibitem [{\citenamefont {Arnaudo}\ \emph {et~al.}(2025)\citenamefont {Arnaudo}, \citenamefont {Carballo},\ and\ \citenamefont {Withers}}]{Arnaudo:2025uos}%
  \BibitemOpen
  \bibfield  {author} {\bibinfo {author} {\bibfnamefont {P.}~\bibnamefont {Arnaudo}}, \bibinfo {author} {\bibfnamefont {J.}~\bibnamefont {Carballo}},\ and\ \bibinfo {author} {\bibfnamefont {B.}~\bibnamefont {Withers}},\ }\bibfield  {title} {\bibinfo {title} {{Beyond quasinormal modes: a complete mode decomposition of black hole perturbations}},\ }\href@noop {} {\  (\bibinfo {year} {2025})},\ \Eprint {https://arxiv.org/abs/2510.18956} {arXiv:2510.18956 [gr-qc]} \BibitemShut {NoStop}%
\bibitem [{\citenamefont {Arnaudo}\ and\ \citenamefont {Withers}(2026{\natexlab{a}})}]{Arnaudo:2026tcy}%
  \BibitemOpen
  \bibfield  {author} {\bibinfo {author} {\bibfnamefont {P.}~\bibnamefont {Arnaudo}}\ and\ \bibinfo {author} {\bibfnamefont {B.}~\bibnamefont {Withers}},\ }\bibfield  {title} {\bibinfo {title} {{Bouncing singularities in Schwarzschild: a geometric origin of the QNM convergence region}},\ }\href@noop {} {\  (\bibinfo {year} {2026}{\natexlab{a}})},\ \Eprint {https://arxiv.org/abs/2605.16489} {arXiv:2605.16489 [gr-qc]} \BibitemShut {NoStop}%
\bibitem [{\citenamefont {Fidkowski}\ \emph {et~al.}(2004)\citenamefont {Fidkowski}, \citenamefont {Hubeny}, \citenamefont {Kleban},\ and\ \citenamefont {Shenker}}]{Fidkowski:2003nf}%
  \BibitemOpen
  \bibfield  {author} {\bibinfo {author} {\bibfnamefont {L.}~\bibnamefont {Fidkowski}}, \bibinfo {author} {\bibfnamefont {V.}~\bibnamefont {Hubeny}}, \bibinfo {author} {\bibfnamefont {M.}~\bibnamefont {Kleban}},\ and\ \bibinfo {author} {\bibfnamefont {S.}~\bibnamefont {Shenker}},\ }\bibfield  {title} {\bibinfo {title} {{The Black hole singularity in AdS / CFT}},\ }\href {https://doi.org/10.1088/1126-6708/2004/02/014} {\bibfield  {journal} {\bibinfo  {journal} {JHEP}\ }\textbf {\bibinfo {volume} {02}},\ \bibinfo {pages} {014}},\ \Eprint {https://arxiv.org/abs/hep-th/0306170} {arXiv:hep-th/0306170} \BibitemShut {NoStop}%
\bibitem [{\citenamefont {Festuccia}\ and\ \citenamefont {Liu}(2006)}]{Festuccia:2005pi}%
  \BibitemOpen
  \bibfield  {author} {\bibinfo {author} {\bibfnamefont {G.}~\bibnamefont {Festuccia}}\ and\ \bibinfo {author} {\bibfnamefont {H.}~\bibnamefont {Liu}},\ }\bibfield  {title} {\bibinfo {title} {{Excursions beyond the horizon: Black hole singularities in Yang-Mills theories. I.}},\ }\href {https://doi.org/10.1088/1126-6708/2006/04/044} {\bibfield  {journal} {\bibinfo  {journal} {JHEP}\ }\textbf {\bibinfo {volume} {04}},\ \bibinfo {pages} {044}},\ \Eprint {https://arxiv.org/abs/hep-th/0506202} {arXiv:hep-th/0506202} \BibitemShut {NoStop}%
\bibitem [{\citenamefont {Parisini}\ \emph {et~al.}(2024)\citenamefont {Parisini}, \citenamefont {Skenderis},\ and\ \citenamefont {Withers}}]{Parisini:2023nbd}%
  \BibitemOpen
  \bibfield  {author} {\bibinfo {author} {\bibfnamefont {E.}~\bibnamefont {Parisini}}, \bibinfo {author} {\bibfnamefont {K.}~\bibnamefont {Skenderis}},\ and\ \bibinfo {author} {\bibfnamefont {B.}~\bibnamefont {Withers}},\ }\bibfield  {title} {\bibinfo {title} {{The ambient space formalism}},\ }\href {https://doi.org/10.1007/JHEP05(2024)296} {\bibfield  {journal} {\bibinfo  {journal} {JHEP}\ }\textbf {\bibinfo {volume} {05}},\ \bibinfo {pages} {296}},\ \Eprint {https://arxiv.org/abs/2312.03820} {arXiv:2312.03820 [hep-th]} \BibitemShut {NoStop}%
\bibitem [{\citenamefont {{\v{C}}eplak}\ \emph {et~al.}(2024)\citenamefont {{\v{C}}eplak}, \citenamefont {Liu}, \citenamefont {Parnachev},\ and\ \citenamefont {Valach}}]{Ceplak:2024bja}%
  \BibitemOpen
  \bibfield  {author} {\bibinfo {author} {\bibfnamefont {N.}~\bibnamefont {{\v{C}}eplak}}, \bibinfo {author} {\bibfnamefont {H.}~\bibnamefont {Liu}}, \bibinfo {author} {\bibfnamefont {A.}~\bibnamefont {Parnachev}},\ and\ \bibinfo {author} {\bibfnamefont {S.}~\bibnamefont {Valach}},\ }\bibfield  {title} {\bibinfo {title} {{Black hole singularity from OPE}},\ }\href {https://doi.org/10.1007/JHEP10(2024)105} {\bibfield  {journal} {\bibinfo  {journal} {JHEP}\ }\textbf {\bibinfo {volume} {10}},\ \bibinfo {pages} {105}},\ \Eprint {https://arxiv.org/abs/2404.17286} {arXiv:2404.17286 [hep-th]} \BibitemShut {NoStop}%
\bibitem [{\citenamefont {Buri{\'c}}\ \emph {et~al.}(2025)\citenamefont {Buri{\'c}}, \citenamefont {Gusev},\ and\ \citenamefont {Parnachev}}]{Buric:2025anb}%
  \BibitemOpen
  \bibfield  {author} {\bibinfo {author} {\bibfnamefont {I.}~\bibnamefont {Buri{\'c}}}, \bibinfo {author} {\bibfnamefont {I.}~\bibnamefont {Gusev}},\ and\ \bibinfo {author} {\bibfnamefont {A.}~\bibnamefont {Parnachev}},\ }\bibfield  {title} {\bibinfo {title} {{Thermal holographic correlators and KMS condition}},\ }\href {https://doi.org/10.1007/JHEP09(2025)053} {\bibfield  {journal} {\bibinfo  {journal} {JHEP}\ }\textbf {\bibinfo {volume} {09}},\ \bibinfo {pages} {053}},\ \Eprint {https://arxiv.org/abs/2505.10277} {arXiv:2505.10277 [hep-th]} \BibitemShut {NoStop}%
\bibitem [{\citenamefont {Buri{\'c}}\ \emph {et~al.}(2026)\citenamefont {Buri{\'c}}, \citenamefont {Gusev},\ and\ \citenamefont {Parnachev}}]{Buric:2025fye}%
  \BibitemOpen
  \bibfield  {author} {\bibinfo {author} {\bibfnamefont {I.}~\bibnamefont {Buri{\'c}}}, \bibinfo {author} {\bibfnamefont {I.}~\bibnamefont {Gusev}},\ and\ \bibinfo {author} {\bibfnamefont {A.}~\bibnamefont {Parnachev}},\ }\bibfield  {title} {\bibinfo {title} {{Holographic correlators from thermal bootstrap}},\ }\href {https://doi.org/10.1007/JHEP05(2026)059} {\bibfield  {journal} {\bibinfo  {journal} {JHEP}\ }\textbf {\bibinfo {volume} {05}},\ \bibinfo {pages} {059}},\ \Eprint {https://arxiv.org/abs/2508.08373} {arXiv:2508.08373 [hep-th]} \BibitemShut {NoStop}%
\bibitem [{\citenamefont {Afkhami-Jeddi}\ \emph {et~al.}(2026)\citenamefont {Afkhami-Jeddi}, \citenamefont {Caron-Huot}, \citenamefont {Chakravarty},\ and\ \citenamefont {Maloney}}]{Afkhami-Jeddi:2025wra}%
  \BibitemOpen
  \bibfield  {author} {\bibinfo {author} {\bibfnamefont {N.}~\bibnamefont {Afkhami-Jeddi}}, \bibinfo {author} {\bibfnamefont {S.}~\bibnamefont {Caron-Huot}}, \bibinfo {author} {\bibfnamefont {J.}~\bibnamefont {Chakravarty}},\ and\ \bibinfo {author} {\bibfnamefont {A.}~\bibnamefont {Maloney}},\ }\bibfield  {title} {\bibinfo {title} {{Imprint of the black hole singularity on thermal two-point functions}},\ }\href {https://doi.org/10.1103/1j3s-fzjl} {\bibfield  {journal} {\bibinfo  {journal} {Phys. Rev. D}\ }\textbf {\bibinfo {volume} {114}},\ \bibinfo {pages} {025005} (\bibinfo {year} {2026})},\ \Eprint {https://arxiv.org/abs/2510.21673} {arXiv:2510.21673 [hep-th]} \BibitemShut {NoStop}%
\bibitem [{\citenamefont {Dodelson}\ \emph {et~al.}(2026)\citenamefont {Dodelson}, \citenamefont {Iossa},\ and\ \citenamefont {Karlsson}}]{Dodelson:2025jff}%
  \BibitemOpen
  \bibfield  {author} {\bibinfo {author} {\bibfnamefont {M.}~\bibnamefont {Dodelson}}, \bibinfo {author} {\bibfnamefont {C.}~\bibnamefont {Iossa}},\ and\ \bibinfo {author} {\bibfnamefont {R.}~\bibnamefont {Karlsson}},\ }\bibfield  {title} {\bibinfo {title} {{Bouncing off a stringy singularity}},\ }\href {https://doi.org/10.1007/JHEP07(2026)161} {\bibfield  {journal} {\bibinfo  {journal} {JHEP}\ }\textbf {\bibinfo {volume} {07}},\ \bibinfo {pages} {161}},\ \Eprint {https://arxiv.org/abs/2511.09616} {arXiv:2511.09616 [hep-th]} \BibitemShut {NoStop}%
\bibitem [{\citenamefont {{\v{C}}eplak}\ \emph {et~al.}(2026)\citenamefont {{\v{C}}eplak}, \citenamefont {Liu}, \citenamefont {Parnachev},\ and\ \citenamefont {Valach}}]{Ceplak:2025dds}%
  \BibitemOpen
  \bibfield  {author} {\bibinfo {author} {\bibfnamefont {N.}~\bibnamefont {{\v{C}}eplak}}, \bibinfo {author} {\bibfnamefont {H.}~\bibnamefont {Liu}}, \bibinfo {author} {\bibfnamefont {A.}~\bibnamefont {Parnachev}},\ and\ \bibinfo {author} {\bibfnamefont {S.}~\bibnamefont {Valach}},\ }\bibfield  {title} {\bibinfo {title} {{Fooling the censor: going beyond inner horizons with the OPE}},\ }\href {https://doi.org/10.1007/JHEP05(2026)001} {\bibfield  {journal} {\bibinfo  {journal} {JHEP}\ }\textbf {\bibinfo {volume} {05}},\ \bibinfo {pages} {001}},\ \Eprint {https://arxiv.org/abs/2511.09638} {arXiv:2511.09638 [hep-th]} \BibitemShut {NoStop}%
\bibitem [{\citenamefont {Jia}\ and\ \citenamefont {Rangamani}(2026{\natexlab{a}})}]{Jia:2025jbi}%
  \BibitemOpen
  \bibfield  {author} {\bibinfo {author} {\bibfnamefont {H.~F.}\ \bibnamefont {Jia}}\ and\ \bibinfo {author} {\bibfnamefont {M.}~\bibnamefont {Rangamani}},\ }\bibfield  {title} {\bibinfo {title} {{Thermal spectral function asymptotics and black hole singularity in holography}},\ }\href {https://doi.org/10.1103/tzm2-jjvg} {\bibfield  {journal} {\bibinfo  {journal} {Phys. Rev. D}\ }\textbf {\bibinfo {volume} {114}},\ \bibinfo {pages} {046025} (\bibinfo {year} {2026}{\natexlab{a}})},\ \Eprint {https://arxiv.org/abs/2512.15114} {arXiv:2512.15114 [hep-th]} \BibitemShut {NoStop}%
\bibitem [{\citenamefont {Giombi}\ \emph {et~al.}(2026)\citenamefont {Giombi}, \citenamefont {Li},\ and\ \citenamefont {Shan}}]{Giombi:2026kdz}%
  \BibitemOpen
  \bibfield  {author} {\bibinfo {author} {\bibfnamefont {S.}~\bibnamefont {Giombi}}, \bibinfo {author} {\bibfnamefont {Y.-Z.}\ \bibnamefont {Li}},\ and\ \bibinfo {author} {\bibfnamefont {J.}~\bibnamefont {Shan}},\ }\bibfield  {title} {\bibinfo {title} {{Bouncing singularities and thermal correlators on line defects}},\ }\href {https://doi.org/10.1007/JHEP07(2026)170} {\bibfield  {journal} {\bibinfo  {journal} {JHEP}\ }\textbf {\bibinfo {volume} {07}},\ \bibinfo {pages} {170}},\ \Eprint {https://arxiv.org/abs/2603.11012} {arXiv:2603.11012 [hep-th]} \BibitemShut {NoStop}%
\bibitem [{\citenamefont {Arnaudo}\ and\ \citenamefont {Withers}(2026{\natexlab{b}})}]{Arnaudo:2026der}%
  \BibitemOpen
  \bibfield  {author} {\bibinfo {author} {\bibfnamefont {P.}~\bibnamefont {Arnaudo}}\ and\ \bibinfo {author} {\bibfnamefont {B.}~\bibnamefont {Withers}},\ }\bibfield  {title} {\bibinfo {title} {{Analytic structure of holographic thermal correlators from Fourier series}},\ }\href {https://doi.org/10.1007/JHEP06(2026)205} {\bibfield  {journal} {\bibinfo  {journal} {JHEP}\ }\textbf {\bibinfo {volume} {06}},\ \bibinfo {pages} {205}},\ \Eprint {https://arxiv.org/abs/2603.13469} {arXiv:2603.13469 [hep-th]} \BibitemShut {NoStop}%
\bibitem [{\citenamefont {Grozdanov}\ \emph {et~al.}(2026{\natexlab{a}})\citenamefont {Grozdanov}, \citenamefont {Valach},\ and\ \citenamefont {Vrbica}}]{Grozdanov:2026cut}%
  \BibitemOpen
  \bibfield  {author} {\bibinfo {author} {\bibfnamefont {S.}~\bibnamefont {Grozdanov}}, \bibinfo {author} {\bibfnamefont {S.}~\bibnamefont {Valach}},\ and\ \bibinfo {author} {\bibfnamefont {M.}~\bibnamefont {Vrbica}},\ }\bibfield  {title} {\bibinfo {title} {{Bouncing geodesics, black hole singularities, and singularities of thermal correlators}},\ }\href {https://doi.org/10.1007/JHEP08(2026)022} {\bibfield  {journal} {\bibinfo  {journal} {JHEP}\ }\textbf {\bibinfo {volume} {08}},\ \bibinfo {pages} {022}},\ \Eprint {https://arxiv.org/abs/2603.15598} {arXiv:2603.15598 [hep-th]} \BibitemShut {NoStop}%
\bibitem [{\citenamefont {Jia}\ and\ \citenamefont {Rangamani}(2026{\natexlab{b}})}]{Jia:2026ryl}%
  \BibitemOpen
  \bibfield  {author} {\bibinfo {author} {\bibfnamefont {H.~F.}\ \bibnamefont {Jia}}\ and\ \bibinfo {author} {\bibfnamefont {M.}~\bibnamefont {Rangamani}},\ }\bibfield  {title} {\bibinfo {title} {{Exact holographic thermal spectral functions: OPE, non-perturbative corrections, and black hole singularity}},\ }\href@noop {} {\  (\bibinfo {year} {2026}{\natexlab{b}})},\ \Eprint {https://arxiv.org/abs/2604.10803} {arXiv:2604.10803 [hep-th]} \BibitemShut {NoStop}%
\bibitem [{\citenamefont {Grozdanov}\ \emph {et~al.}(2026{\natexlab{b}})\citenamefont {Grozdanov}, \citenamefont {Movrin},\ and\ \citenamefont {Valach}}]{Grozdanov:2026ktq}%
  \BibitemOpen
  \bibfield  {author} {\bibinfo {author} {\bibfnamefont {S.}~\bibnamefont {Grozdanov}}, \bibinfo {author} {\bibfnamefont {V.}~\bibnamefont {Movrin}},\ and\ \bibinfo {author} {\bibfnamefont {S.}~\bibnamefont {Valach}},\ }\bibfield  {title} {\bibinfo {title} {{Bouncing Geodesics, Singularities, and the Cavity Thermal Product Formula in Asymptotically Flat and de Sitter Black Holes}},\ }\href@noop {} {\  (\bibinfo {year} {2026}{\natexlab{b}})},\ \Eprint {https://arxiv.org/abs/2606.11297} {arXiv:2606.11297 [hep-th]} \BibitemShut {NoStop}%
\bibitem [{\citenamefont {Arnaudo}\ \emph {et~al.}(2026)\citenamefont {Arnaudo}, \citenamefont {Iossa}, \citenamefont {Karlsson},\ and\ \citenamefont {Withers}}]{Arnaudo:2026axe}%
  \BibitemOpen
  \bibfield  {author} {\bibinfo {author} {\bibfnamefont {P.}~\bibnamefont {Arnaudo}}, \bibinfo {author} {\bibfnamefont {C.}~\bibnamefont {Iossa}}, \bibinfo {author} {\bibfnamefont {R.}~\bibnamefont {Karlsson}},\ and\ \bibinfo {author} {\bibfnamefont {B.}~\bibnamefont {Withers}},\ }\bibfield  {title} {\bibinfo {title} {{OPE = QNM}},\ }\href@noop {} {\  (\bibinfo {year} {2026})},\ \Eprint {https://arxiv.org/abs/2607.24909} {arXiv:2607.24909 [hep-th]} \BibitemShut {NoStop}%
\bibitem [{\citenamefont {Barrat}\ \emph {et~al.}(2026)\citenamefont {Barrat}, \citenamefont {Bozkurt}, \citenamefont {Marchetto}, \citenamefont {Miscioscia},\ and\ \citenamefont {Pomoni}}]{Barrat:2026jfg}%
  \BibitemOpen
  \bibfield  {author} {\bibinfo {author} {\bibfnamefont {J.}~\bibnamefont {Barrat}}, \bibinfo {author} {\bibfnamefont {D.~N.}\ \bibnamefont {Bozkurt}}, \bibinfo {author} {\bibfnamefont {E.}~\bibnamefont {Marchetto}}, \bibinfo {author} {\bibfnamefont {A.}~\bibnamefont {Miscioscia}},\ and\ \bibinfo {author} {\bibfnamefont {E.}~\bibnamefont {Pomoni}},\ }\bibfield  {title} {\bibinfo {title} {{Analytic thermal bootstrap in momentum space: From thermal OPE to QNMs}},\ }\href@noop {} {\  (\bibinfo {year} {2026})},\ \Eprint {https://arxiv.org/abs/2607.24919} {arXiv:2607.24919 [hep-th]} \BibitemShut {NoStop}%
\bibitem [{\citenamefont {Berti}\ \emph {et~al.}(2006)\citenamefont {Berti}, \citenamefont {Cardoso},\ and\ \citenamefont {Casals}}]{Berti:2005gp}%
  \BibitemOpen
  \bibfield  {author} {\bibinfo {author} {\bibfnamefont {E.}~\bibnamefont {Berti}}, \bibinfo {author} {\bibfnamefont {V.}~\bibnamefont {Cardoso}},\ and\ \bibinfo {author} {\bibfnamefont {M.}~\bibnamefont {Casals}},\ }\bibfield  {title} {\bibinfo {title} {{Eigenvalues and eigenfunctions of spin-weighted spheroidal harmonics in four and higher dimensions}},\ }\href {https://doi.org/10.1103/PhysRevD.73.109902} {\bibfield  {journal} {\bibinfo  {journal} {Phys. Rev. D}\ }\textbf {\bibinfo {volume} {73}},\ \bibinfo {pages} {024013} (\bibinfo {year} {2006})},\ \bibinfo {note} {[Erratum: Phys.Rev.D 73, 109902 (2006)]},\ \Eprint {https://arxiv.org/abs/gr-qc/0511111} {arXiv:gr-qc/0511111} \BibitemShut {NoStop}%
\bibitem [{\citenamefont {Keshet}\ and\ \citenamefont {Hod}(2007)}]{Keshet:2007nv}%
  \BibitemOpen
  \bibfield  {author} {\bibinfo {author} {\bibfnamefont {U.}~\bibnamefont {Keshet}}\ and\ \bibinfo {author} {\bibfnamefont {S.}~\bibnamefont {Hod}},\ }\bibfield  {title} {\bibinfo {title} {{Analytic study of rotating black-hole quasinormal modes}},\ }\href {https://doi.org/10.1103/PhysRevD.76.061501} {\bibfield  {journal} {\bibinfo  {journal} {Phys. Rev. D}\ }\textbf {\bibinfo {volume} {76}},\ \bibinfo {pages} {061501} (\bibinfo {year} {2007})},\ \Eprint {https://arxiv.org/abs/0705.1179} {arXiv:0705.1179 [gr-qc]} \BibitemShut {NoStop}%
\bibitem [{\citenamefont {Keshet}\ and\ \citenamefont {Ben-Meir}(2014)}]{Keshet:2012uq}%
  \BibitemOpen
  \bibfield  {author} {\bibinfo {author} {\bibfnamefont {U.}~\bibnamefont {Keshet}}\ and\ \bibinfo {author} {\bibfnamefont {A.}~\bibnamefont {Ben-Meir}},\ }\bibfield  {title} {\bibinfo {title} {{The missing asymptotic sector of rotating black-hole spectroscopy}},\ }\href {https://doi.org/10.1016/j.physletb.2014.09.011} {\bibfield  {journal} {\bibinfo  {journal} {Phys. Lett. B}\ }\textbf {\bibinfo {volume} {737}},\ \bibinfo {pages} {379} (\bibinfo {year} {2014})},\ \Eprint {https://arxiv.org/abs/1207.2460} {arXiv:1207.2460 [gr-qc]} \BibitemShut {NoStop}%
\bibitem [{\citenamefont {Teukolsky}(1972)}]{PhysRevLett.29.1114}%
  \BibitemOpen
  \bibfield  {author} {\bibinfo {author} {\bibfnamefont {S.~A.}\ \bibnamefont {Teukolsky}},\ }\bibfield  {title} {\bibinfo {title} {Rotating black holes: Separable wave equations for gravitational and electromagnetic perturbations},\ }\href {https://doi.org/10.1103/PhysRevLett.29.1114} {\bibfield  {journal} {\bibinfo  {journal} {Phys. Rev. Lett.}\ }\textbf {\bibinfo {volume} {29}},\ \bibinfo {pages} {1114} (\bibinfo {year} {1972})}\BibitemShut {NoStop}%
\bibitem [{\citenamefont {Oshita}\ \emph {et~al.}(2025)\citenamefont {Oshita}, \citenamefont {Ma}, \citenamefont {Chen},\ and\ \citenamefont {Yang}}]{Oshita:2025qmn}%
  \BibitemOpen
  \bibfield  {author} {\bibinfo {author} {\bibfnamefont {N.}~\bibnamefont {Oshita}}, \bibinfo {author} {\bibfnamefont {S.}~\bibnamefont {Ma}}, \bibinfo {author} {\bibfnamefont {Y.}~\bibnamefont {Chen}},\ and\ \bibinfo {author} {\bibfnamefont {H.}~\bibnamefont {Yang}},\ }\bibfield  {title} {\bibinfo {title} {{Probing Direct Waves in Black Hole Ringdowns}},\ }\href@noop {} {\  (\bibinfo {year} {2025})},\ \Eprint {https://arxiv.org/abs/2509.09165} {arXiv:2509.09165 [gr-qc]} \BibitemShut {NoStop}%
\bibitem [{\citenamefont {Su}\ \emph {et~al.}(2026)\citenamefont {Su}, \citenamefont {Khera}, \citenamefont {Chowdhuri}, \citenamefont {Casals},\ and\ \citenamefont {Yang}}]{Su:2026gmp}%
  \BibitemOpen
  \bibfield  {author} {\bibinfo {author} {\bibfnamefont {J.}~\bibnamefont {Su}}, \bibinfo {author} {\bibfnamefont {N.}~\bibnamefont {Khera}}, \bibinfo {author} {\bibfnamefont {A.}~\bibnamefont {Chowdhuri}}, \bibinfo {author} {\bibfnamefont {M.}~\bibnamefont {Casals}},\ and\ \bibinfo {author} {\bibfnamefont {H.}~\bibnamefont {Yang}},\ }\bibfield  {title} {\bibinfo {title} {{Gravitational Waves from Green's Function Decomposition for a Kerr black hole: I. Equatorial ISCO Plunge}},\ }\href@noop {} {\  (\bibinfo {year} {2026})},\ \Eprint {https://arxiv.org/abs/2608.17943} {arXiv:2608.17943 [gr-qc]} \BibitemShut {NoStop}%
\end{thebibliography}%

\end{document}